\documentclass[useAMS,usenatbib,usegraphicx]{mn2e}
\usepackage{psfig}   
\usepackage{graphicx}
\usepackage{amssymb}
\usepackage{subfigure}

\title[Binaries in cool, clumpy star clusters]{The evolution of binary populations in cool, clumpy star clusters}

\author[R.~J.~Parker, S.~P.~Goodwin and R.~J.~Allison]{
  Richard J.~Parker$^1$\thanks{E-mail: rparker@phys.ethz.ch},
  Simon P.~Goodwin$^2$ and Richard J.~Allison$^3$ \vspace*{0.1cm}\\
   $^1$ Institute for Astronomy, ETH Z{\"u}rich, Wolfgang-Pauli-Strasse 27, 8093, Z{\"u}rich, Switzerland \\
   $^2$ Department of Physics and Astronomy, University of Sheffield,
    Sheffield, S3 7RH, UK\\
   $^3$ Zentrum f{\"u}r Astronomie der Universit{\"a}t Heidelberg, Institut f{\"u}r Theoretische Astrophysik, Albert-Ueberle-Str. 2, 69120 Heidelberg, Germany}

\begin{document}

\date{Accepted for publication in MNRAS}
                             
\pagerange{\pageref{firstpage}--\pageref{lastpage}} \pubyear{2011}

\maketitle

\label{firstpage}

\begin{abstract}
Observations and theory suggest that star clusters can form in a subvirial (cool) state and are highly substructured. Such initial conditions have been proposed to explain 
the level of mass segregation in clusters through dynamics, and have also been successful in explaining the origin of trapezium-like systems. In this 
paper we investigate, using $N$-body simulations, whether such a dynamical scenario is consistent with the observed binary properties in the Orion Nebula Cluster (ONC). We find that 
several different primordial binary populations are consistent with the overall fraction and separation distribution of visual binaries in the ONC (in the range 67 -- 670\,au), and that these binary systems 
are heavily processed. The substructured, cool-collapse scenario requires a primordial binary fraction approaching 100 per cent. We find that the most important factor in 
processing the primordial binaries is the initial level of substructure; a highly substructured cluster processes up to 20 per cent more systems than a less substructured cluster because of localised pockets of high stellar density 
in the substructure. Binaries are processed in the substructure before the cluster reaches its densest phase, suggesting that even clusters remaining in virial equilibrium or undergoing supervirial expansion would 
dynamically alter their primordial binary population. Therefore even some expanding associations may not preserve their primordial binary population.  
\end{abstract}

\begin{keywords}   
stars: formation -- kinematics and dynamics -- open clusters and associations: general -- methods: numerical
\end{keywords}

\section{Introduction}

It is thought that the vast majority of stars form in clustered environments \citep*[with surface densities of several stars, to several hundred stars per square parsec, e.g.][]{Lada03,Lada10,Zwart10}. Whether all such clusters are dense enough to dynamically process 
the primordial stellar population is currently the subject of debate \citep[e.g.][]{Bressert10}. However, there is observational and theoretical evidence that some clusters do at 
least undergo a dense phase in their evolution, a notable example being the Orion Nebula Cluster (ONC).

Recent work by \citet{Allison09b} has shown that the observed mass segregation in the ONC can be of a dynamical origin. If a cluster is initially substructured 
(\citet{Allison09b} used fractals to create substructure) and subvirial, then the cluster undergoes cool-collapse and the most massive stars mass segregate, 
in some cases forming trapezium-like systems \citep{Allison11}. Previously, it had been thought that the mass segregation in the ONC had to be primordial 
\citep{Bonnell98}, as the level of dynamical mass segregation required cannot occur within $\sim$~1\,Myr in clusters with smooth radial profiles.

Given the success of the cool-collapse model in producing the observed levels of mass segregation and trapezium systems, an investigation into the effects of this 
dynamical scenario on clusters containing primordial binary populations is timely. For simplicity, \citet{Allison09b} did not include primordial binaries in their simulations. However, 
the binary fraction in the ONC is consistent with that in the field \citep[$\sim$ 45 per cent,][]{Petr98,Reipurth07}. $N$-body simulations 
by \citet*{Kroupa95a,Kroupa95b,Kroupa99}, and more recently, \citet{Parker09a}, have shown that in a dense cluster in virial equilibrium, a binary population with a high primordial binary 
fraction ($\sim$ 100 per cent) will be processed to a much lower binary fraction, consistent with the observations in the ONC. \citet{Parker09a} argued that it was unlikely 
that the primordial binary population in the ONC was field-like, as the ONC is expanding (indicating that it was much denser in the past and therefore had a higher primordial binary fraction) and there are no wide ($>$1000\,au) binary systems \citep*{Scally99} (indicating 
that the binary population has been well processed). 

In this paper, we investigate the effect of dynamical evolution in substructured, subvirial clusters on various primordial binary populations. We run suites of $N$-body simulations 
in which we vary the initial amount of substructure, and the proportion of stars in binary systems, and compare the results to the most recent observations of binaries in the ONC. 
In Section~\ref{method} we describe the set-up of the clusters, and the initial binary populations; we present our results in Section~\ref{results}; we provide a discussion in Section~\ref{discuss}, 
and we conclude in Section~\ref{conclude}.  

\section{Method}
\label{method}

\subsection{Initial conditions}

The clusters we simulate have 1500 members, which corresponds to a cluster mass of $\sim 10^3$ M$_\odot$. For each set of initial conditions, 
we run an ensemble of 10 simulations, identical apart from the random number seed used to initialise the positions, masses and binary properties. 

Our clusters are set up as fractals; observations of young unevolved star forming regions indicate a high level of substructure is present \citep[i.e. they do not  
have a radially smooth profile, e.g.][and references therein]{Cartwright04,Sanchez09,Schmeja11}. The fractal distribution provides a way of creating substructure on all scales. 
Note that we are not claiming that young star clusters are fractal \citep[although they may be, e.g.][]{Elmegreen01}, but the fractal distribution is a relatively simple method of 
setting up substructured clusters, as the level of substructure is described by just one parameter, the fractal dimension, $D$. In three dimensions, $D = 1.6$ indicates a 
highly substructured cluster, and $D = 3.0$ is a roughly uniform sphere.

We set up the fractals according to the method in \citet{Goodwin04a}. This begins by defining a cube of side $N_{\rm div}$ (we adopt $N_{\rm div} = 2.0$ 
throughout), inside of which the fractal is built. A first-generation parent is placed at the centre of the cube, which then spawns $N_{\rm div}^3$ subcubes, each containing a first generation 
child at its centre. The fractal is then built by determining which of the children themselves become parents, and spawn their own offspring. This is determined by the 
fractal dimension, $D$, where the probability that the child becomes a parent is given by $N_{\rm div}^{(D - 3)}$. For a lower fractal dimension fewer children 
mature and the final distribution contains more substructure. Any children that do not become parents in a given step are removed, along with all of their parents. 
A small amount of noise is then 
added to the positions of the remaining children, preventing the cluster from having a gridded appearance and the children become parents of the next generation. Each new parent 
then spawns $N_{\rm div}^3$ second-generation children in $N_{\rm div}^3$ sub-subcubes, with each second-generation child having a $N_{\rm div}^{(D - 3)}$ 
probability of becoming a second generation parent. This process is repeated until there are substantially more children than required. The children are pruned to produce a 
sphere from the cube and are then randomly removed (so maintaining the fractal dimension) until the required number of children is left. These children then become stars in the 
cluster. 

To determine the velocity structure of the cloud, children inherit their parent's velocity plus a random component that decreases with each generation of the fractal.  The children of the first 
generation are given random velocities from a Gaussian of mean zero. Each new generation inherits their parent's velocity plus an extra random component that becomes smaller with each 
generation. This results in a velocity structure in which nearby stars have similar velocities, but distant stars can have very different velocities. The velocity of every star is scaled to obtain the desired virial 
ratio of the cluster.

We set up clusters with fractal dimensions of $D = 1.6$ (very clumpy), $D = 2.0$ and $D = 3.0$ (a roughly uniform sphere), in order to investigate 
the full parameter space.  The clusters are out of virial equilibrium at the start of the simulations and have a virial ratio of $Q = 0.3$, where we define the virial ratio as $Q = T/|\Omega|$ ($T$ and $|\Omega|$ 
are the total kinetic energy and total potential energy of the stars, respectively). Therefore, a cluster with $Q = 0.5$ is in virial equilibrium and a cluster with $Q = 0.3$ is `subvirial', or `cool'. 

To create a stellar system, the mass of the primary star is chosen randomly from a \citet{Kroupa02} IMF of the form
\begin{equation}
 N(M)   \propto  \left\{ \begin{array}{ll} M^{-1.3} \hspace{0.4cm} m_0
  < M/{\rm M_\odot} \leq m_1   \,, \\ M^{-2.3} \hspace{0.4cm} m_1 <
  M/{\rm M_\odot} \leq m_2   \,,
\end{array} \right.
\end{equation}
where $m_0$ = 0.1\,M$_\odot$, $m_1$ = 0.5\,M$_\odot$, and  $m_2$ =
50\,M$_\odot$.  We do not include brown dwarfs in the simulations; the binary properties of brown dwarfs and very low mass stars appear to be very different from those of M-, K-, and G-dwarfs 
\citep[e.g.][]{Burgasser07,Thies08}. For a fuller discussion of the effects of dynamical processing on brown dwarfs we refer the interested reader to the work of \citet{Kroupa03} and \citet{Parker11}. 

We then assign a secondary component to the system depending on the binary fraction associated with the primary mass. For a field-like binary fraction we divide primaries into four groups. 
 Primary masses in the range 0.1~$\leq M/{\rm M}_\odot~<$~0.47 are M-dwarfs, with a binary fraction of 0.42 \citep{Fischer92}. K-dwarfs have masses in  the range 
 0.47~$\leq~M/{\rm M}_\odot$~$<$~0.84 with a binary fraction of 0.45 \citep{Mayor92}, and G-dwarfs have masses from 0.84~$\leq~M/{\rm M}_\odot~<$~1.2  with a binary fraction of 0.57 
\citep{Duquennoy91,Raghavan10}. All stars more massive than  1.2\,M$_\odot$ are grouped together and assigned a binary fraction of unity, as massive stars have a much larger binary fraction than low-mass stars
\citep[e.g.][and references therein]{Abt90,Mason98,Kouwenhoven05,Kouwenhoven07,Pfalzner07,Mason09}.

We also set up clusters with a binary fraction of unity for all stars, and a binary fraction of 0.75 for all stars, according to the hypothesis that most, if not all stars, form in binary systems and that single stars are purely the result of 
dynamical processing of binaries and higher-order systems \citep{Kroupa95a,Goodwin05a}.

\subsection{Binary properties}

Secondary masses are drawn from a flat mass ratio distribution; recent work by \citet{Reggiani11} has shown the companion mass ratio of field stars to be consistent with being drawn from a flat 
distribution, rather than random pairing from the IMF. Currently, however, there is no detailed statistical analysis for the ONC. We note that drawing companions from a flat distribution means we do not recover 
a Kroupa IMF.

We draw the periods of the binary systems from two generating functions. Firstly, in accordance with observations of the field, we use the log$_{10}$-normal fit to the G-dwarfs in the field by \citet[][hereafter DM91]{Duquennoy91} -- see also \citet{Raghavan10}, which has also been extrapolated to fit the period distributions of the K- and M-dwarfs \citep{Mayor92,Fischer92}:
\begin{equation}
f\left({\rm log_{10}}P\right)  \propto {\rm exp}\left \{ \frac{-{({\rm log_{10}}P -
\overline{{\rm log_{10}}P})}^2}{2\sigma^2_{{\rm log_{10}}P}}\right \},
\end{equation}
where $\overline{{\rm log_{10}}P} = 4.8$, $\sigma_{{\rm log_{10}}P} = 2.3$ and $P$ is  in days. Alternatively, we draw periods from the initial pre-main sequence period function derived by \citet[][hereafter K95]{Kroupa95a,Kroupa95b}: 
\begin{equation}
f\left({\rm log_{10}}P\right) = \eta\frac{{\rm log_{10}}P - {\rm log_{10}}P_{\rm min}}{\delta + \left({\rm log_{10}}P - {\rm log_{10}}P_{\rm min}\right)^2},
\end{equation}
where ${\rm log_{10}}P_{\rm min}$ is the logarithm of the minimum period in days. We adopt ${\rm log_{10}}P_{\rm min} = 0$; and $\eta = 3.5$ and $\delta = 100$ are the numerical 
constants adopted by \citet{Kroupa95a} and \citet{Kroupa11} to fit the observed pre-main sequence distributions.  We convert the periods to semi-major axes using the masses of the binary components.

The eccentricities of binary stars are drawn from a thermal distribution \citep{Heggie75,Kroupa08} of the form
\begin{equation}
f_e(e) = 2e.
\end{equation}
In the sample of \citet{Duquennoy91}, close binaries (with periods less than 10 days) are almost exclusively on tidally circularised orbits. We account for this by reselecting the eccentricity of a system if it exceeds the following 
period-dependent value\footnote{\citet{Kroupa95b} and \citet{Kroupa08} provides a more elaborate `eigenevolution' mechanism to incorporate interactions between the primary star and its protostellar disk during tidal circularisation. 
However, this mechanism also alters the mass ratio distribution, causing a deviation from the flat mass ratio distribution observed in the Galactic field \citep{Reggiani11}.}:
\begin{equation}
e_{\rm tid} = \frac{1}{2}\left[0.95 + {\rm tanh}\left(0.6\,{\rm log_{10}}P - 1.7\right)\right].
\end{equation}

We combine the primary and secondary masses of the binaries with their semi-major axes and eccentricities to determine the relative velocity and radial components of the stars in each system. 
The binaries are then placed at the centre of mass and velocity for each system in the fractal. The simulations are run for 10\,Myr using the \texttt{kira} integrator in the Starlab package 
\citep[e.g.][]{Zwart99,Zwart01}. We do not include stellar evolution in the simulations. Details of each simulation are presented in Table~\ref{cluster_props}.

\begin{table}
\caption[bf]{A summary of the different cluster properties adopted for the simulations.
The values in the columns are: the number of stars in each cluster ($N_{\rm stars}$), 
the typical mass of this cluster ($M_{\rm cluster}$), the initial fractal dimension of 
the cluster ($D$), the initial binary fraction in the cluster ($f_{\rm bin}$), 
and the primordial binary separation distribution (either \citet[][DM91]{Duquennoy91} 
or \citet[][K95]{Kroupa95a}).}
\begin{center}
\begin{tabular}{|c|c|c|c|c|}
\hline 
$N_{\rm stars}$ & $M_{\rm cluster}$  & $D$ &  $f_{\rm bin}$ & Separations \\
\hline
1500 & $\sim 10^3$\,M$_\odot$ & 1.6  & 100\,per cent & DM91 \\
1500 & $\sim 10^3$\,M$_\odot$ & 1.6  & 100\,per cent & K95 \\
1500 & $\sim 10^3$\,M$_\odot$ & 1.6  & field-like & DM91 \\
1500 & $\sim 10^3$\,M$_\odot$ & 1.6  & 75\,per cent & DM91 \\
\hline 
1500 & $\sim 10^3$\,M$_\odot$ & 2.0  & 100\,per cent & DM91 \\
1500 & $\sim 10^3$\,M$_\odot$ & 2.0  & 100\,per cent & K95 \\
1500 & $\sim 10^3$\,M$_\odot$ & 2.0  & field-like & DM91 \\
1500 & $\sim 10^3$\,M$_\odot$ & 2.0  & 75\,per cent & DM91 \\
\hline
1500 & $\sim 10^3$\,M$_\odot$ & 3.0  & 100\,per cent & DM91 \\
1500 & $\sim 10^3$\,M$_\odot$ & 3.0  & 100\,per cent & K95 \\
1500 & $\sim 10^3$\,M$_\odot$ & 3.0  & field-like & DM91 \\
1500 & $\sim 10^3$\,M$_\odot$ & 3.0  & 75\,per cent & DM91 \\
\hline
\end{tabular}
\end{center}
\label{cluster_props}
\end{table}

\section{Results}
\label{results}

In this section we will describe the results of dynamical evolution on the primordial binary population in subvirial clusters with three differing levels of substructure, as set by the fractal 
dimension. We consider clusters with an initial fractal dimensions of $D = 1.6$ (highly substructured)  $D = 2.0$, and $D = 3.0$ (almost no initial substructure). We first examine the evolution of the 
substructure in the clusters, before following the evolution of the binary populations by looking at the overall binary fractions and separation distributions. 

We determine whether a star is in a bound binary system using the nearest-neighbour method outlined in \citet{Parker09a} and \citet{Kouwenhoven10}.

\subsection{Cluster morphologies and evolution}

\begin{figure*}
  \begin{center}
\setlength{\subfigcapskip}{10pt}
\subfigure[$D = 1.6$, 0 Myr]{\label{frac1p6_i}\rotatebox{270}{\includegraphics[scale=0.35]{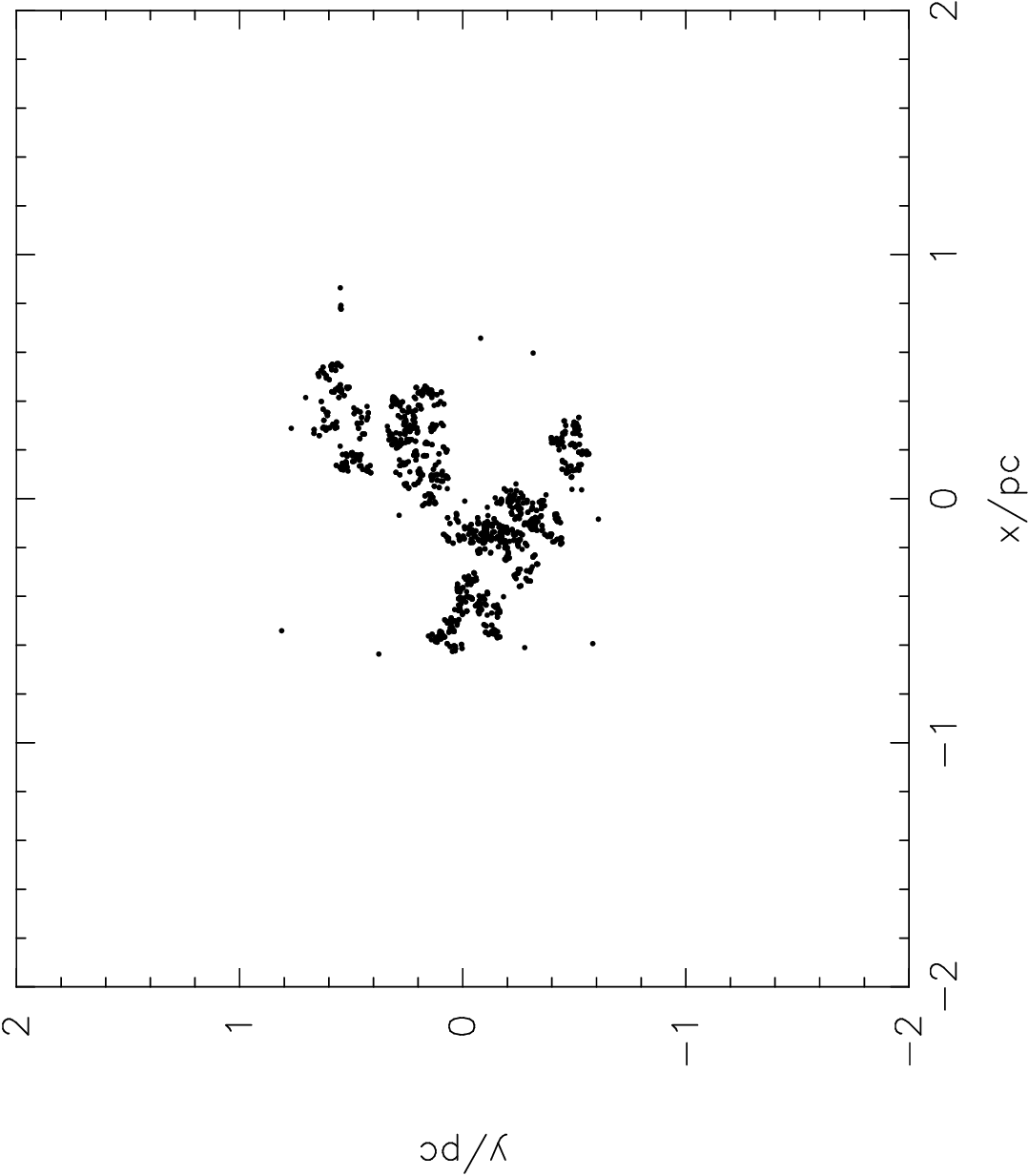}}}
\hspace*{0.6cm}
\subfigure[$D = 1.6$, 1 Myr]{\label{frac1p6_s}\rotatebox{270}{\includegraphics[scale=0.35]{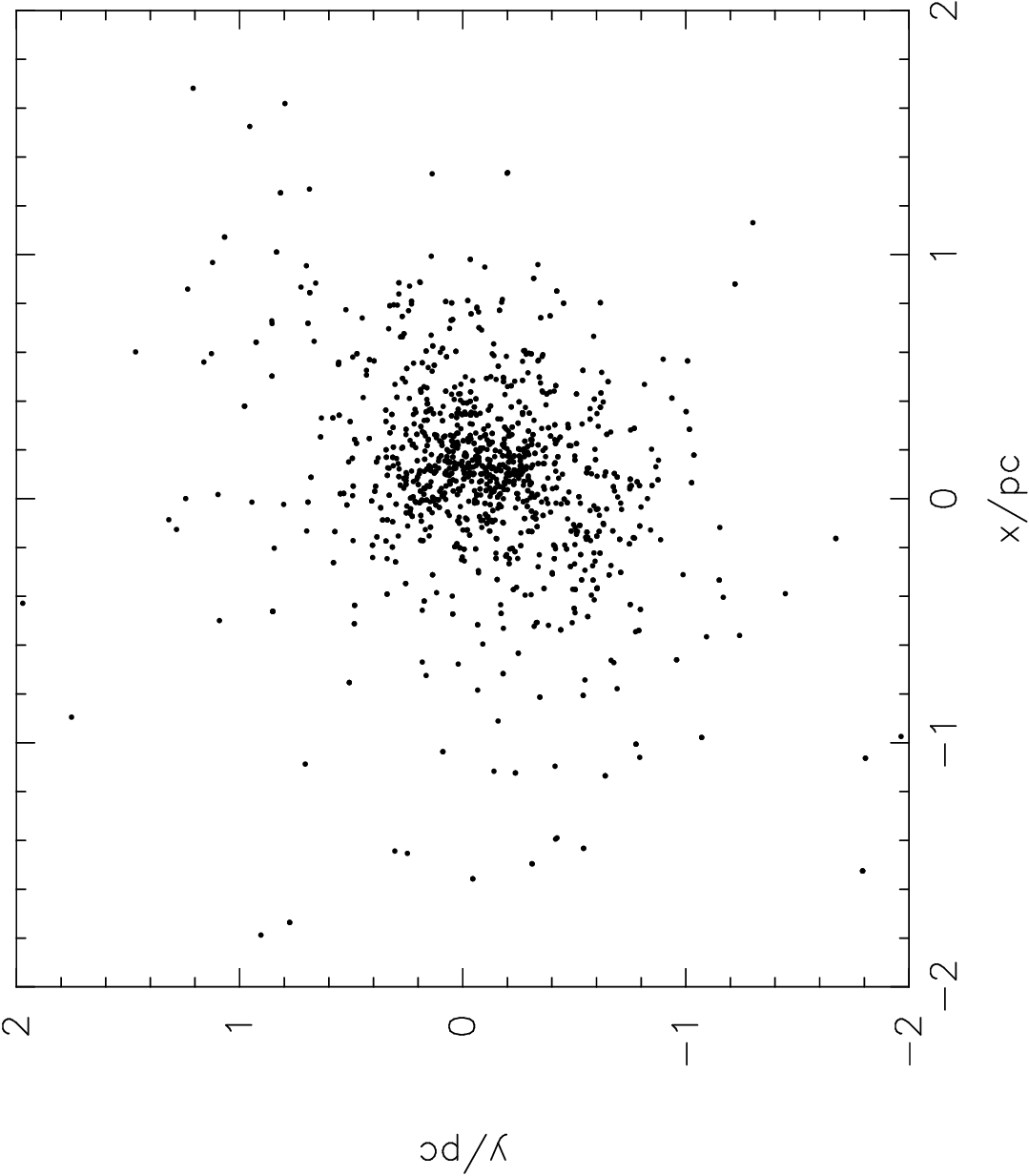}}}
\subfigure[$D = 2.0$, 0 Myr]{\label{frac2p0_i}\rotatebox{270}{\includegraphics[scale=0.35]{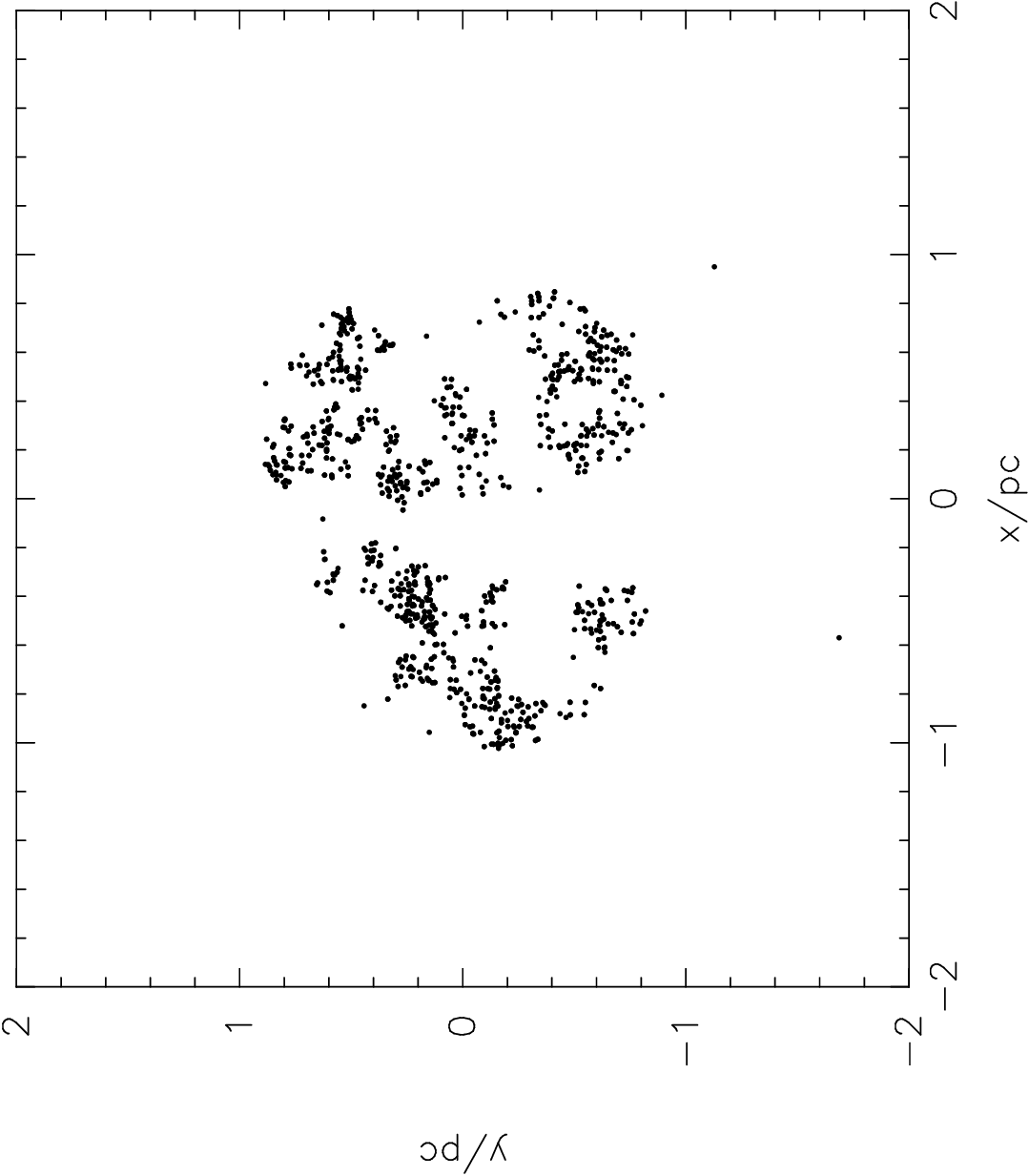}}}
\hspace*{0.6cm}
\subfigure[$D = 2.0$, 1 Myr]{\label{frac2p0_s}\rotatebox{270}{\includegraphics[scale=0.35]{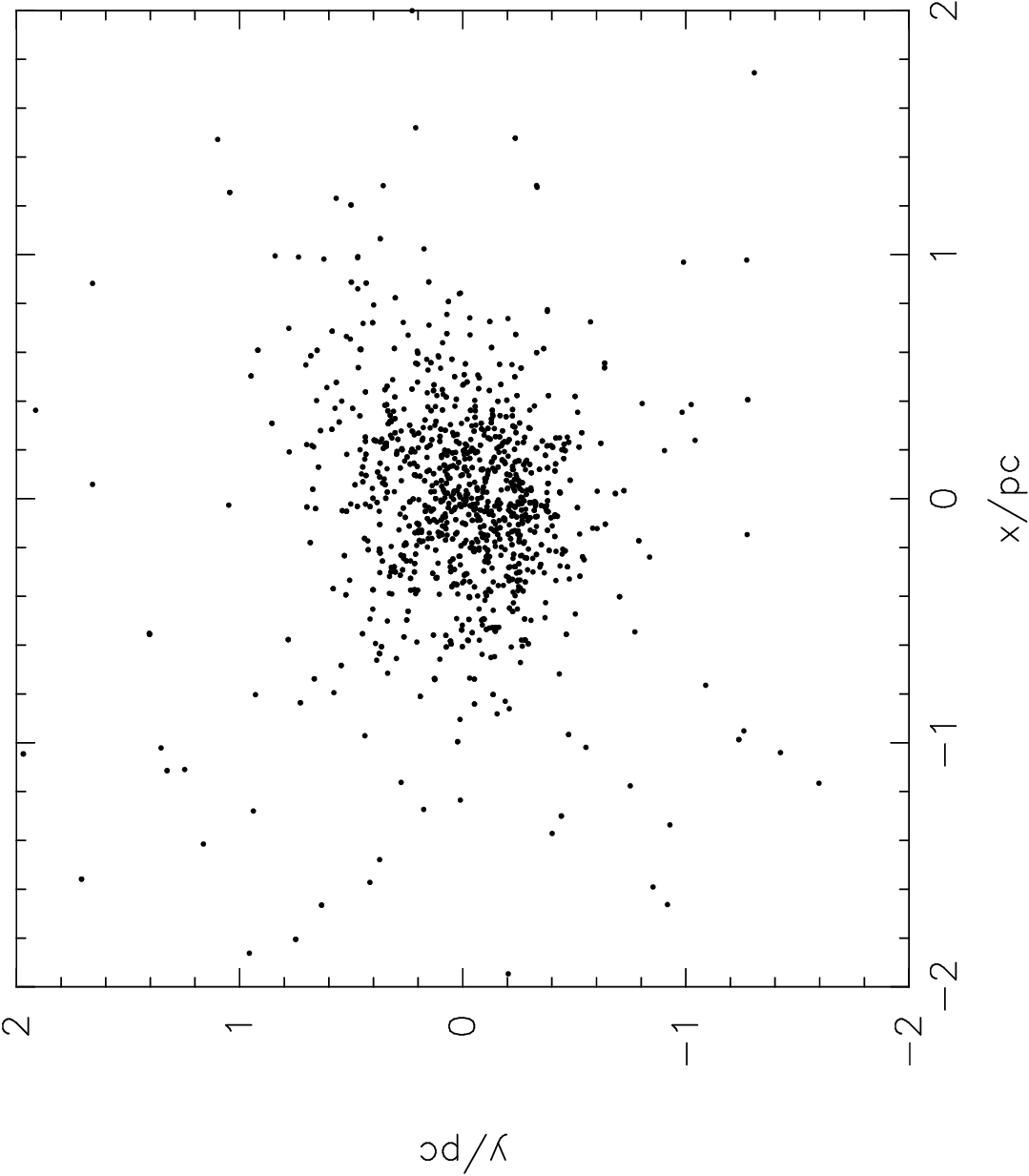}}}
\vspace*{0.25cm}
\subfigure[$D = 3.0$, 0 Myr]{\label{frac3p0_i}\rotatebox{270}{\includegraphics[scale=0.35]{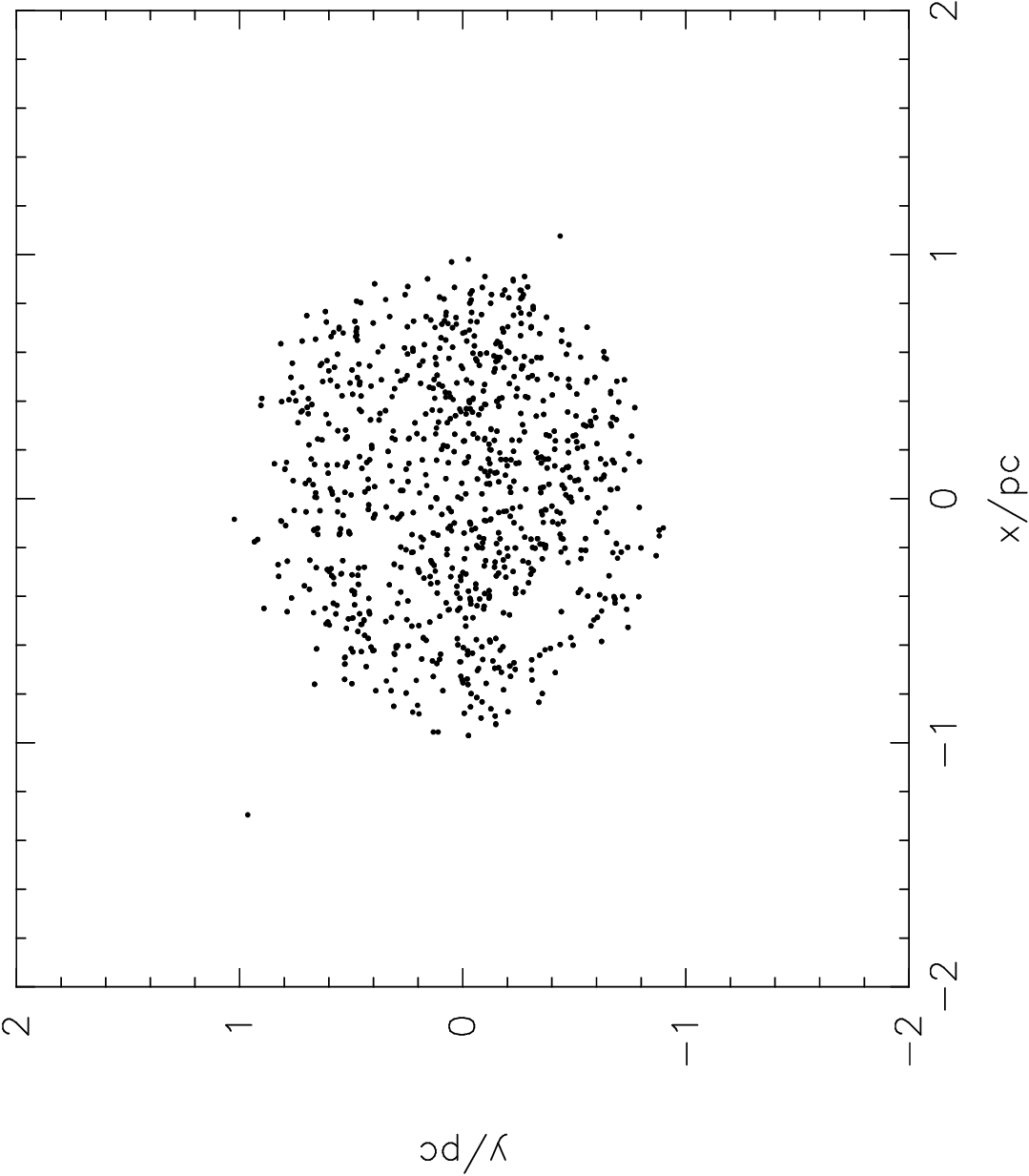}}}
\hspace*{0.6cm}
\subfigure[$D = 3.0$, 1 Myr]{\label{frac3p0_s}\rotatebox{270}{\includegraphics[scale=0.35]{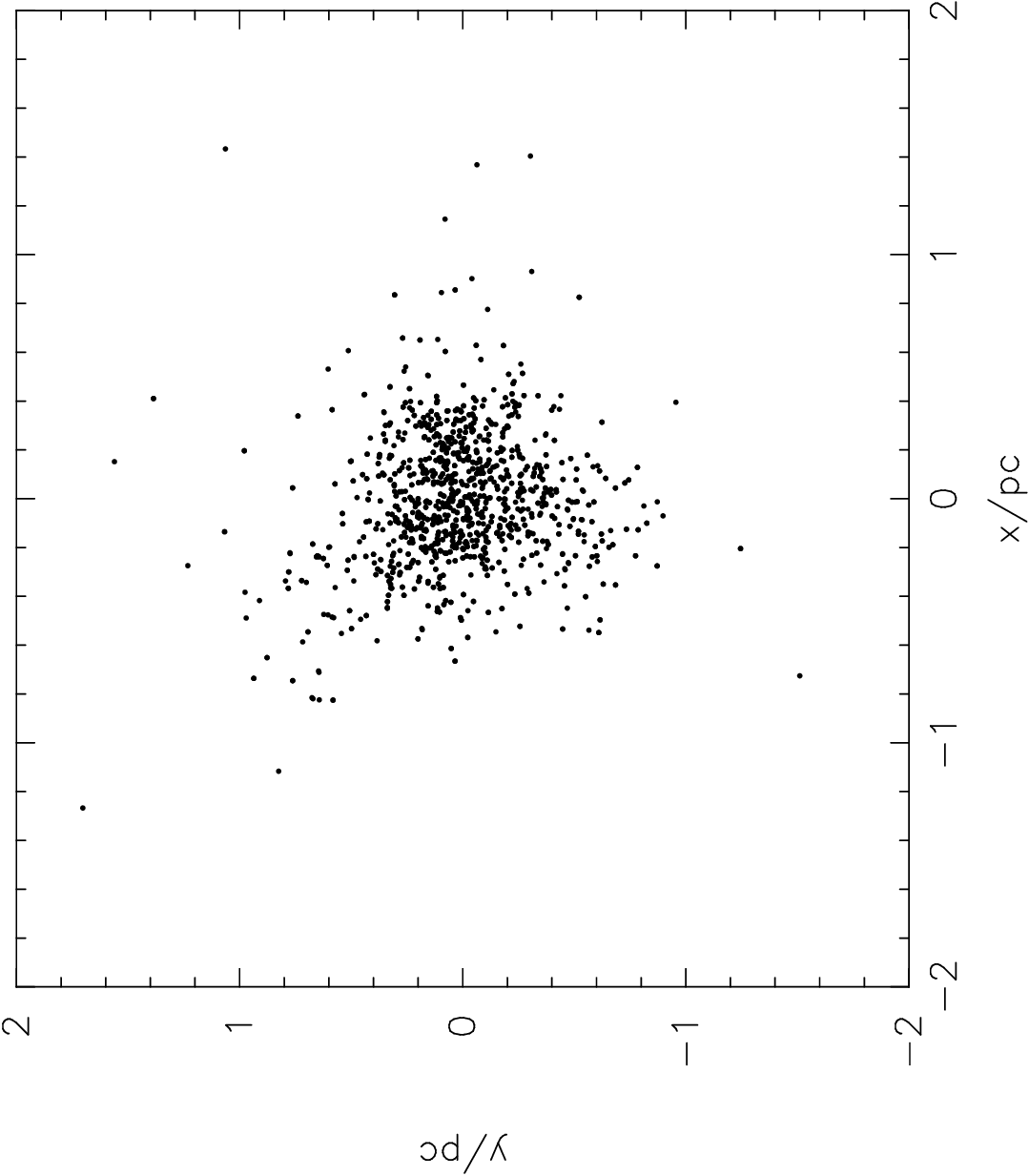}}}
  \end{center}
  \caption[bf]{Morphologies of typical examples of our clusters with an initial fractal dimension $D = 1.6$ at (a) 0\,Myr and (b) 1\,Myr;   $D = 2.0$ at (c) 0\,Myr and (d) 1\,Myr; and  $D = 3.0$ at (e) 0\,Myr and (f) 1\,Myr.}
  \label{morph_stars}
\end{figure*}

In Fig.~\ref{morph_stars} we show typical examples of initial cluster morphologies for the three initial levels of substructure. As found by \citet{Allison09b,Allison10}, the clusters collapse on very 
short timescales ($< 1$\,Myr), leading to Plummer sphere-like morphologies on timescales of order the age of the ONC \citep[1\,Myr;][]{Jeffries07a,Jeffries07b}. Irrespective of the initial fractal dimension, the 
clusters reach similar morphologies after 1\,Myr. 

\begin{figure}
\rotatebox{270}{\includegraphics[scale=0.35]{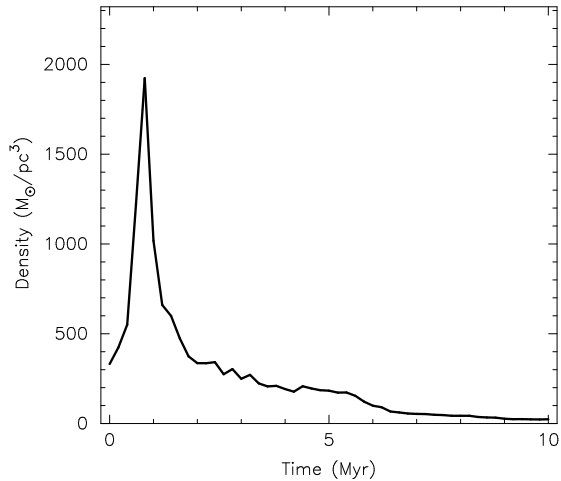}}
 \caption[bf]{The core density of a cluster with fractal dimension $D = 2.0$, undergoing cool collapse, as a function of time. The cluster reaches a peak density of 1920\,M$_\odot$\,pc$^{-3}$  at 0.9\,Myr.}
  \label{cluster_evolve}
\end{figure}

In Fig.~\ref{cluster_evolve} we show the evolution of a typical cluster `core'\footnote{Note that it is meaningless to define a `core' for a fractal cluster before it undergoes collapse. Before the collapse and formation of a core, we calculate the density within the half-mass radius from the centre of mass of the cluster.} 
over the lifetime of the simulation. We have picked a $D = 2.0$ simulation, but clusters with different fractal dimensions exhibit very similar 
behaviour. In this figure we plot the core density of the cluster as a function of time. The initial density of the fractal is 330\,M$_\odot$\,pc$^{-3}$, which increases to 1920\,M$_\odot$\,pc$^{-3}$ during the densest phase at 0.9\,Myr, immediately after cool-collapse. 
Following this dense phase the cluster quickly relaxes and after 10\,Myr has a density of 25\,M$_\odot$\,pc$^{-3}$. 

\subsection{Evolution of the binary fraction}

\begin{figure*}
  \begin{center}
\setlength{\subfigcapskip}{10pt}
\hspace*{-0.3cm}
\subfigure[$D = 1.6$]{\label{binfrac1p6}\rotatebox{270}{\includegraphics[scale=0.27]{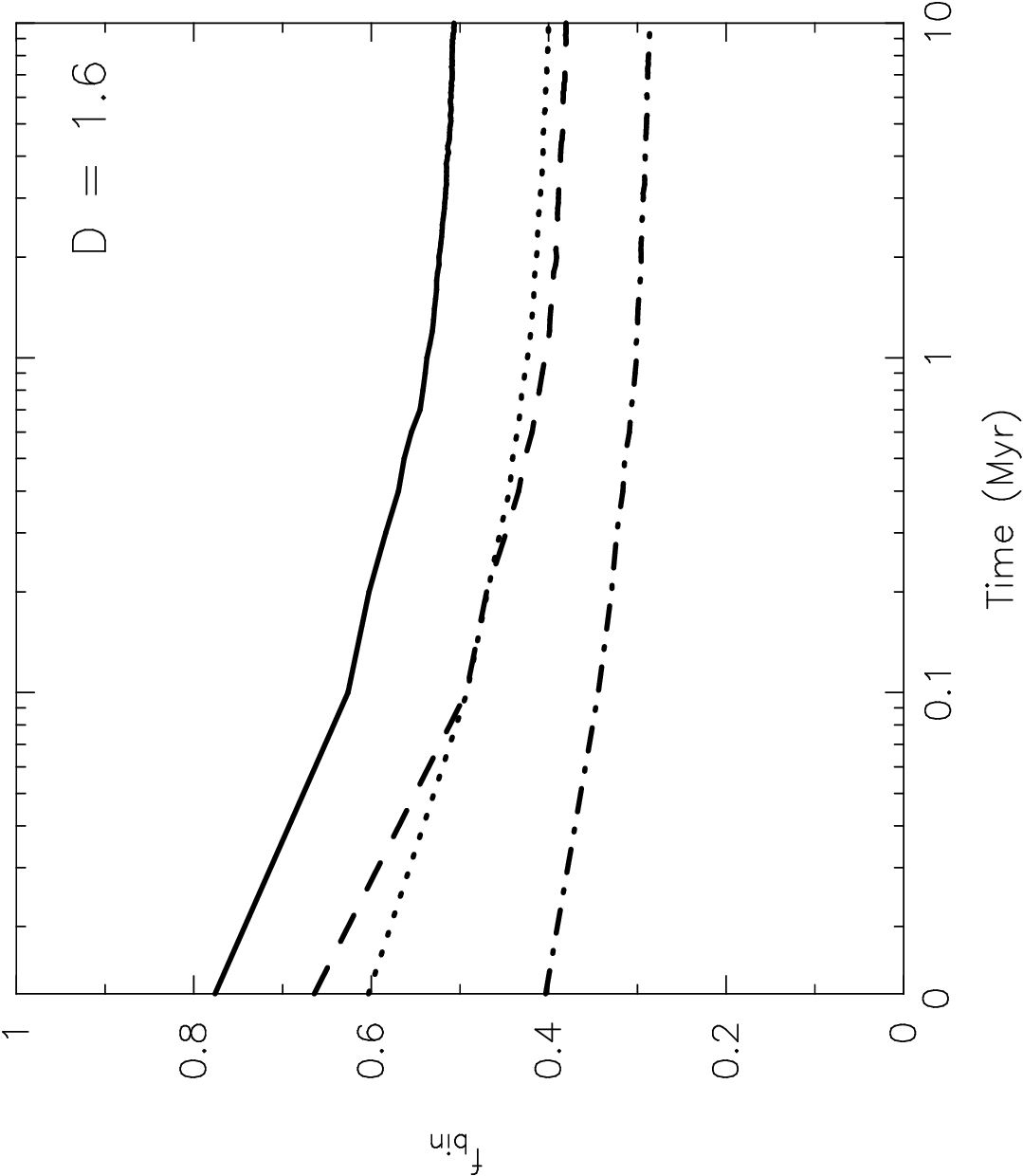}}}
\hspace*{0.2cm}
\subfigure[$D = 2.0$]{\label{binfrac2p0}\rotatebox{270}{\includegraphics[scale=0.27]{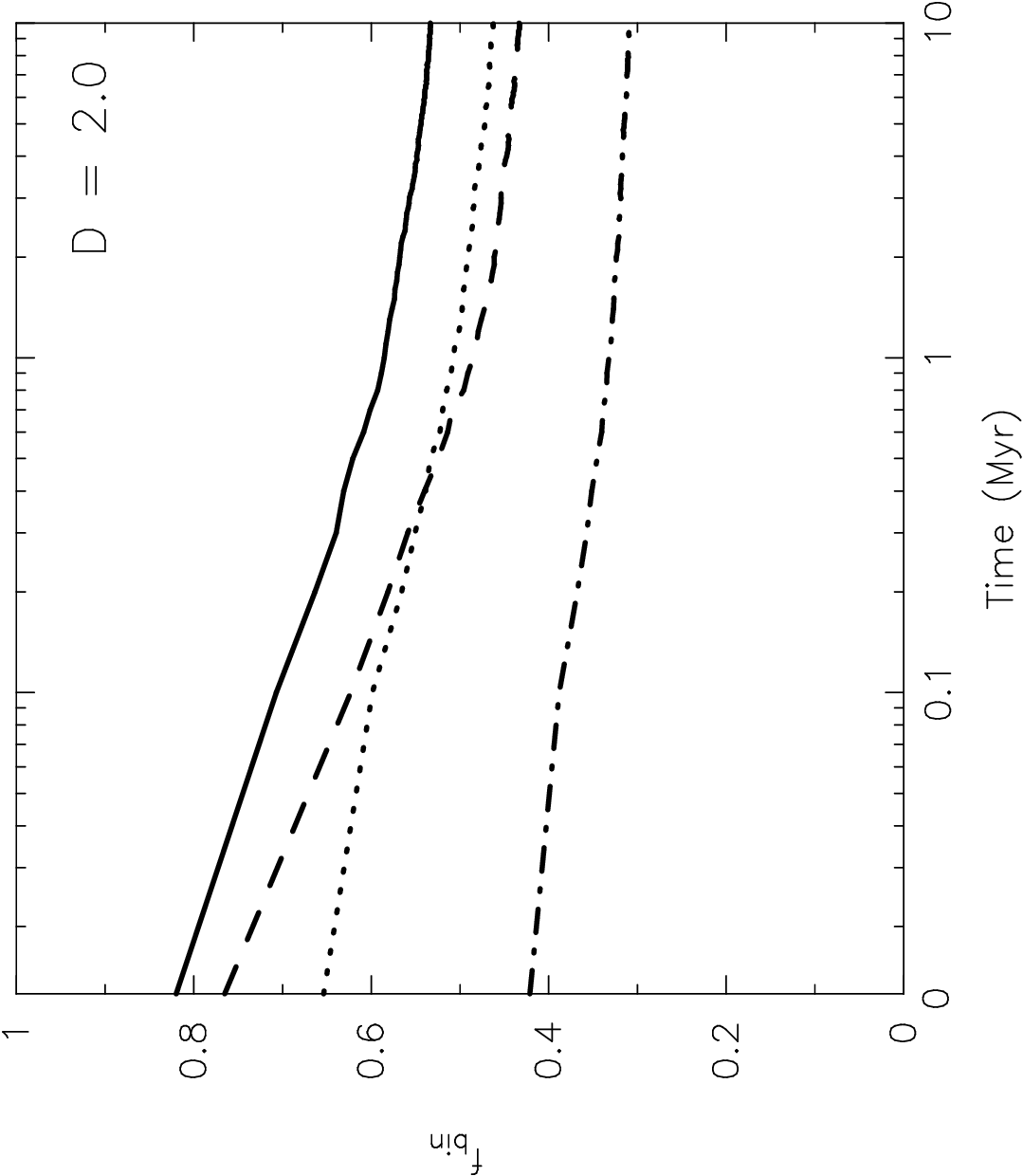}}}
\hspace*{0.2cm}
\subfigure[$D = 3.0$]{\label{binfrac3p0}\rotatebox{270}{\includegraphics[scale=0.27]{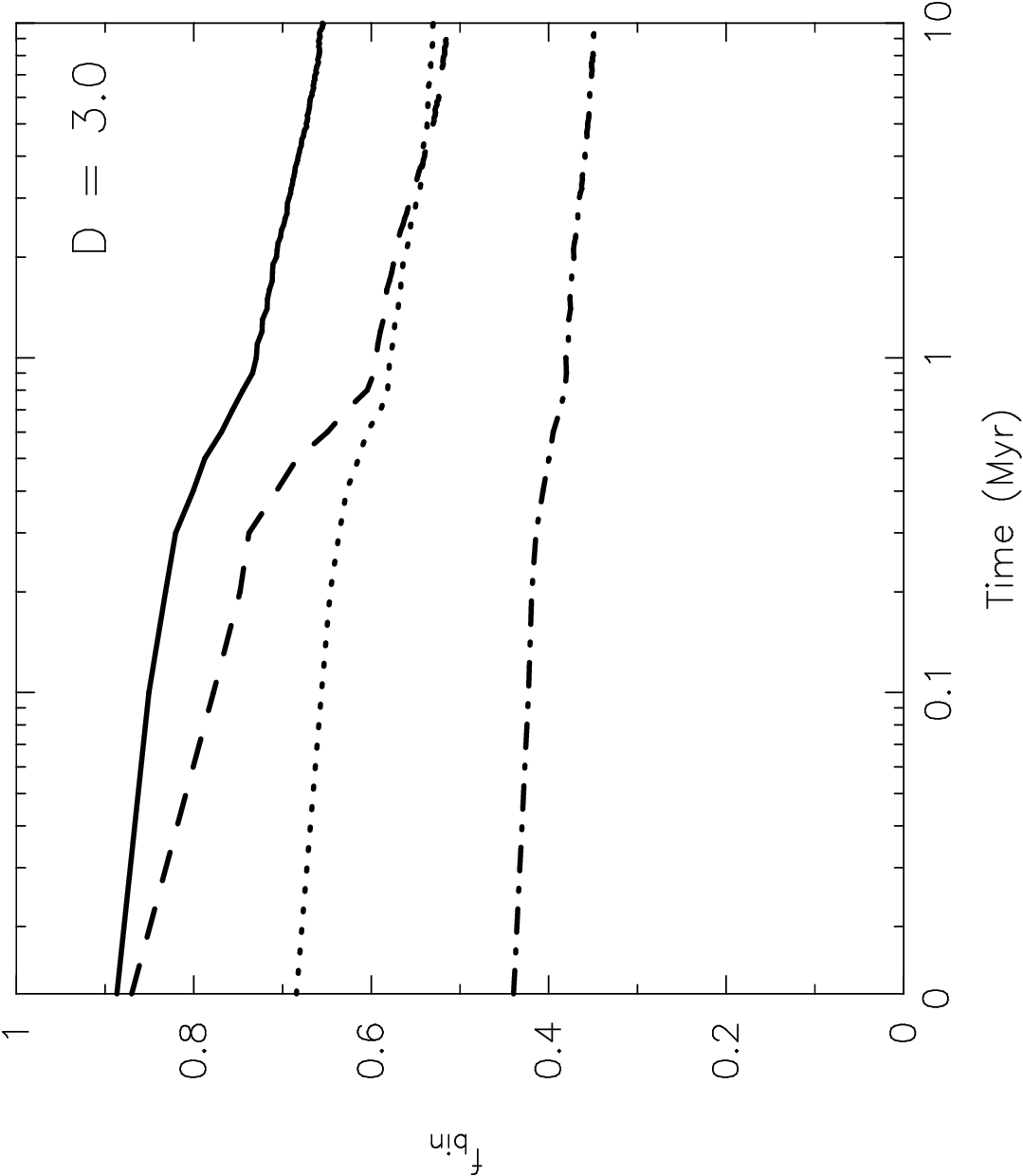}}}
  \end{center}
  \caption[bf]{The evolution of the binary fraction in clusters with different amounts of substructure; (a) a very clumpy cluster (fractal dimension $D = 1.6$), (b) a moderately substructured cluster ($D = 2.0$) and (c) a roughly uniform sphere ($D = 3.0$). 
Four different primordial binary populations are shown; (i) an initially 100 per cent binary fraction with the 
DM91 separation distribution (the solid line), (ii) an initially 100 per cent binary fraction with the K95 pre-main sequence separation distribution (the dashed line), 
(iii) a DM91 separation distribution with an initially field-like binary fraction (the dot-dashed line), and (iv) a DM91 separation distribution with an initially 75 per cent binary fraction (the dotted line).}
  \label{bin_fracs}
\end{figure*}

In Fig.~\ref{bin_fracs} we show the evolution of the binary fraction over 10\,Myr, averaging together 10 clusters with the same initial fractal dimensions. We show the evolution of the binary fraction for four 
different primordial binary populations; the DM91 distribution with an initially 100 per cent binary fraction (the solid line), the  K95 separation 
distribution with an initially 100 per cent binary fraction (the dashed line), and the DM91 separation distributions with field-like and 75 per cent initial binary fractions (the dot-dashed and dotted lines, respectively).

The results shown in Fig.~\ref{bin_fracs} are summarised in Table~\ref{bin_frac_table}. For the various initial conditions, we show the binary fraction as measured by our algorithm at 0\,Myr, 1\,Myr and 10\,Myr.
When considering the evolution of the binary fraction in dense, virialised Plummer spheres, \citet{Parker09a} noted that the cluster was too dense initially for the widest binaries observed in the field to be bound systems. This means 
that the initial binary fraction in the clusters in \citet{Parker09a} with a 100 per cent primordial binary fraction actually translated into an initial value of 75 per cent for a DM91 separation distribution. The fractal clusters presented here 
are less dense than these Plummer spheres initially, and the calculated binary fractions are all higher (although none are 100 per cent).

For clusters with a moderate level of substructure ($D = 2.0$), an initial input binary fraction of 100 per cent, and separations drawn from the DM91 distribution, the measured binary fraction at 0\,Myr is 83 per cent, higher than
the 75 per cent initial binary fraction in a dense Plummer sphere \citep{Parker09a}.

\begin{table}
\caption[bf]{A summary of the results presented in Fig.~\ref{bin_fracs}. From left to right, the fractal dimension of the cluster ($D$), binary separation distribution ($f\left({\rm log_{10}}P\right)$), initial binary fraction inputted into the simulations 
($f_{\rm bin, init}$), the initial binary fraction as measured by our algorithm ($f_{\rm bin, 0\,Myr}$), the binary fraction after 1\,Myr ($f_{\rm bin, 1\,Myr}$), and the binary fraction after 10\,Myr ($f_{\rm bin, 10\,Myr}$). }
\begin{center}
\begin{tabular}{|c|c|c|c|c|c|}
\hline 
$D$ & $f\left({\rm log_{10}}P\right)$   & $f_{\rm bin, init}$ & $f_{\rm bin, 0\,Myr}$ & $f_{\rm bin, 1\,Myr}$ & $f_{\rm bin, 10\,Myr}$ \\
\hline
1.6 & DM91 & 1.00 & 0.79 & 0.55 & 0.51 \\
1.6 & K95 & 1.00 & 0.68 & 0.40 & 0.38 \\
1.6 & DM91 & 0.75 & 0.62 & 0.44 & 0.40 \\
1.6 & DM91 & 0.45 & 0.41 & 0.30 & 0.28 \\
\hline 
2.0 & DM91 & 1.00 & 0.83 & 0.58 & 0.54 \\
2.0 & K95 & 1.00 & 0.78 & 0.48 & 0.43 \\
2.0 & DM91 & 0.75 & 0.66 & 0.51 & 0.46 \\
2.0 & DM91 & 0.45 & 0.42 & 0.34 & 0.31 \\
\hline
3.0 & DM91 & 1.00 & 0.89 & 0.74 & 0.66 \\
3.0 & K95 & 1.00 & 0.88 & 0.58 & 0.51 \\
3.0 & DM91 & 0.75 & 0.68 & 0.57 & 0.53 \\
3.0 & DM91 & 0.45 & 0.44 & 0.38 & 0.35 \\
\hline
\end{tabular}
\end{center}
\label{bin_frac_table}
\end{table}

This effect is even more pronounced for the clusters with a binary fraction of 100 per cent and separations drawn from the K95 distribution. This distribution was derived to reconcile the observed overabundance of wide binaries in young clusters with the 
DM91 field distribution. Recently, \citet*{Marks11} have suggested that a dynamical operator (which is a function of the cluster's density) can be used to transform a K95 distribution to the field distribution in a dense cluster. However, the K95 distribution saturates 
a dense cluster with wide binaries which are not physically bound, and it is difficult to see how they could form in such an environment. \citet{Marks11} suggest this problem could be negated if the cluster formed in a more sparse environment, and then 
underwent cool collapse, which is exactly the scenario we propose here. However, the calculated initial binary fraction in all the clusters here is significantly lower than 100 per cent (the dashed lines in Fig.~\ref{bin_fracs}; see also Table~\ref{bin_frac_table}), 
which indicates that very wide binaries cannot form in star forming regions; an alternative solution is that they form during cluster dissolution, when two stars are simultaneously ejected in the same direction \citep[e.g.][]{Kouwenhoven10,Moeckel10}.

The initial and final binary fractions depend heavily on the level of substructure. Comparing the simulations with $D = 1.6$ (highly substructured), to those with $D = 3.0$ (uniform spheres), we see that the initial binary fraction is higher by 10 per cent 
for the uniform sphere (0.89 versus 0.79), and after 10\,Myr the difference is still significant, with a binary fraction of 0.66 for the $D = 3.0$ model versus 0.51 for $D = 1.6$ model. 

Indeed, comparison of Figs.~\ref{binfrac1p6}~and~\ref{binfrac2p0} with the overall evolution of the cluster in Fig.~\ref{cluster_evolve} shows that the vast majority of binary processing occurs before the cluster reaches its densest phase 
(after 0.9\,Myr). This is due to pockets of localised density in the substructure, which dynamically process the binary populations. In the case of a cluster with almost no initial substructure (Fig.~\ref{binfrac3p0}), we see that there is very little binary processing until the cluster has almost reached its densest phase at collapse 
(note the sudden drop in binary fraction between 0.3 and 0.9\,Myr, which corresponds to the density peak in Fig.~\ref{cluster_evolve}). 

The fact that dense substructure processes binaries to almost the same extent as the overall collapse of the cluster suggests that substructured clusters in virial equilibrium and those undergoing expansion would also process any primordial binary population.


\citet{Petr98} and \citet{Reipurth07}  estimate that the binary fraction in the ONC is consistent with the field value, i.e. between 40 and 60 per cent, depending on the spectral type of the primary. We note from Fig.~\ref{bin_fracs} 
and Table~\ref{bin_frac_table} that dynamical processing reduces the overall binary fraction to such an extent that the initial binary fraction cannot be that of the field. Even in the smooth clusters, the binary fraction 
at 1\,Myr is less than 40 per cent. For the other initial conditions ($D = 1.6$ or $D = 2.0$), the binary fraction in the ONC at 1\,Myr can be reproduced (within the uncertainties) if the initial binary fraction was 75 per cent or higher. 


\subsection{The complete binary separation distribution}

\begin{figure*}
  \begin{center}
\setlength{\subfigcapskip}{10pt} \subfigure[100 per cent binary fraction, DM91 separation distribution]{\label{sepdist2p0-a}\rotatebox{270}{\includegraphics[scale=0.35]{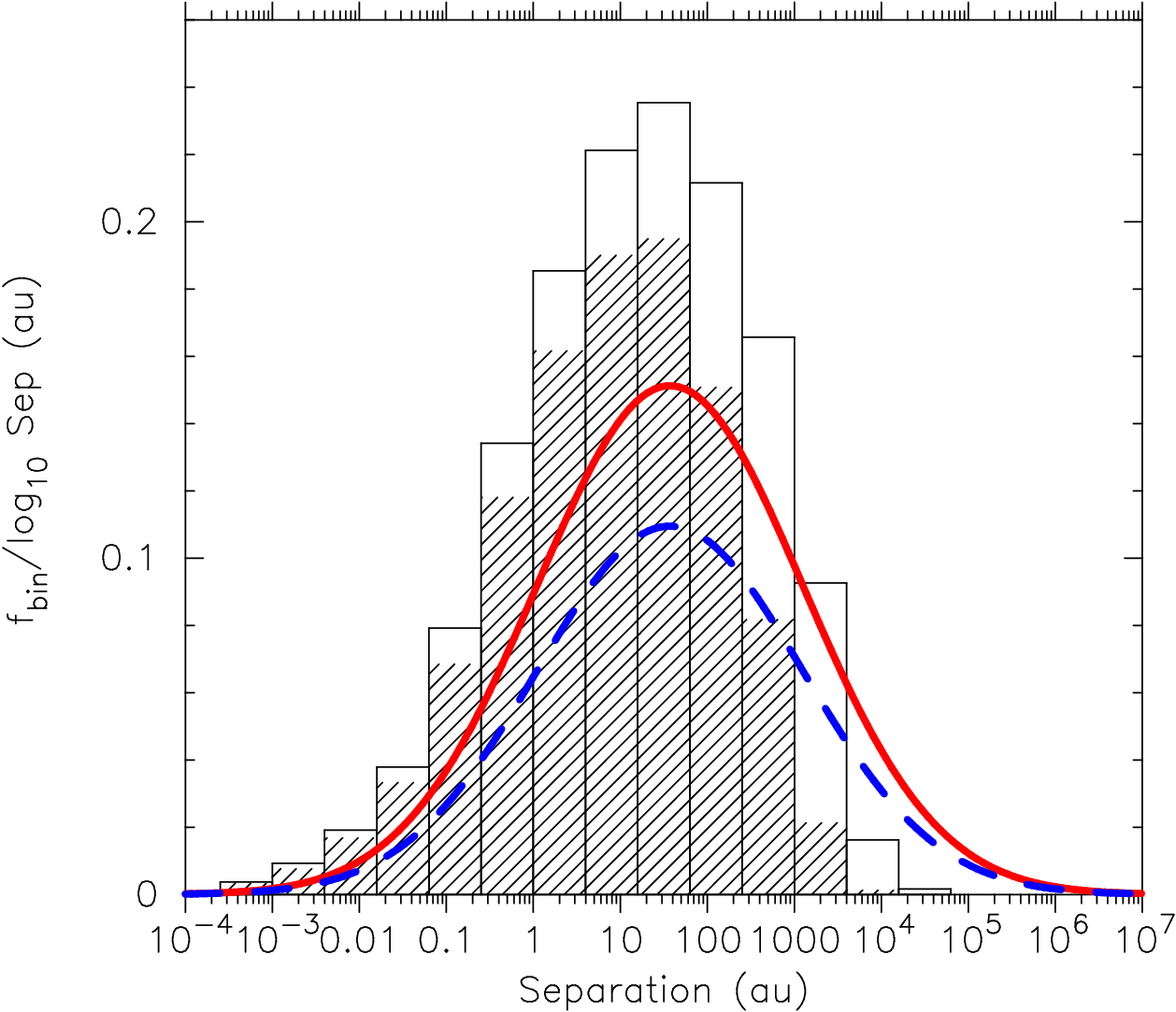}}}
\hspace*{0.6cm} 
\subfigure[100 per cent binary fraction, K95 separation distribution]{\label{sepdist2p0-b}\rotatebox{270}{\includegraphics [scale=0.35]{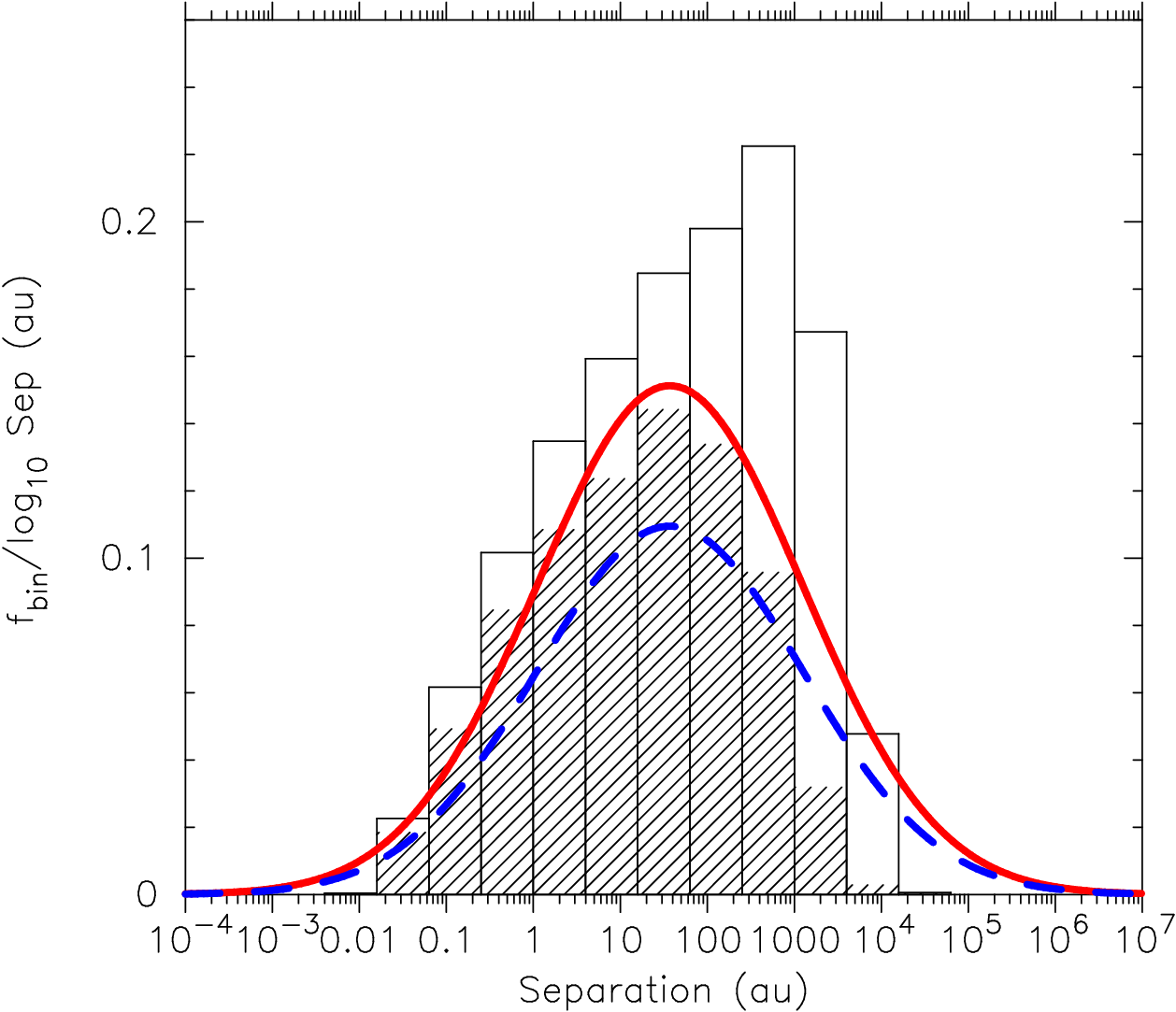}}}
\vspace*{0.5cm} \subfigure[Field binary fraction, DM91 separation distribution]{\label{sepdist2p0-c}\rotatebox{270}{\includegraphics[scale=0.35]{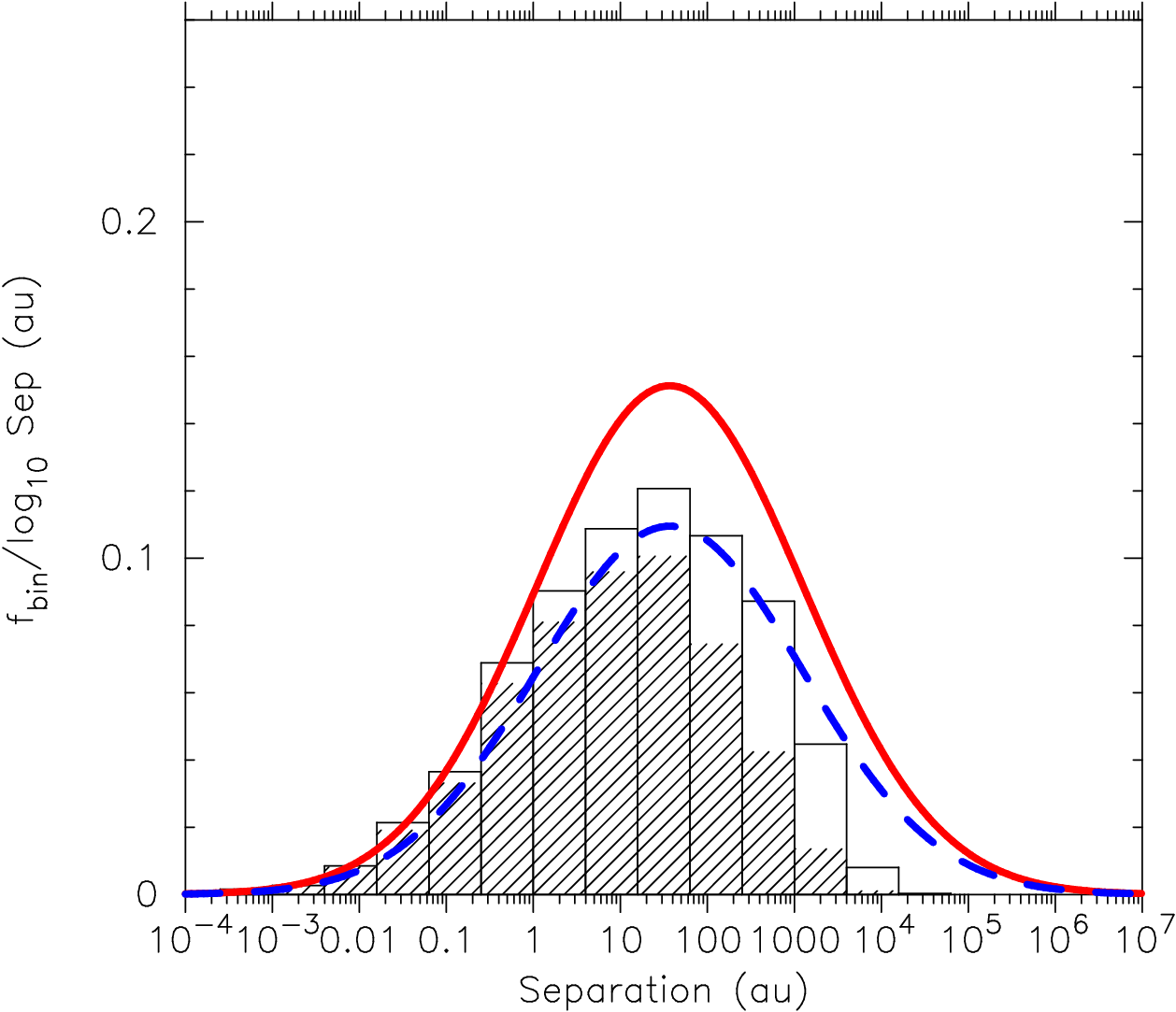}}}
\hspace*{0.6cm} \subfigure[75 per cent binary fraction, DM91 separation distribution]{\label{sepdist2p0-d}\rotatebox{270}{\includegraphics[scale=0.35]{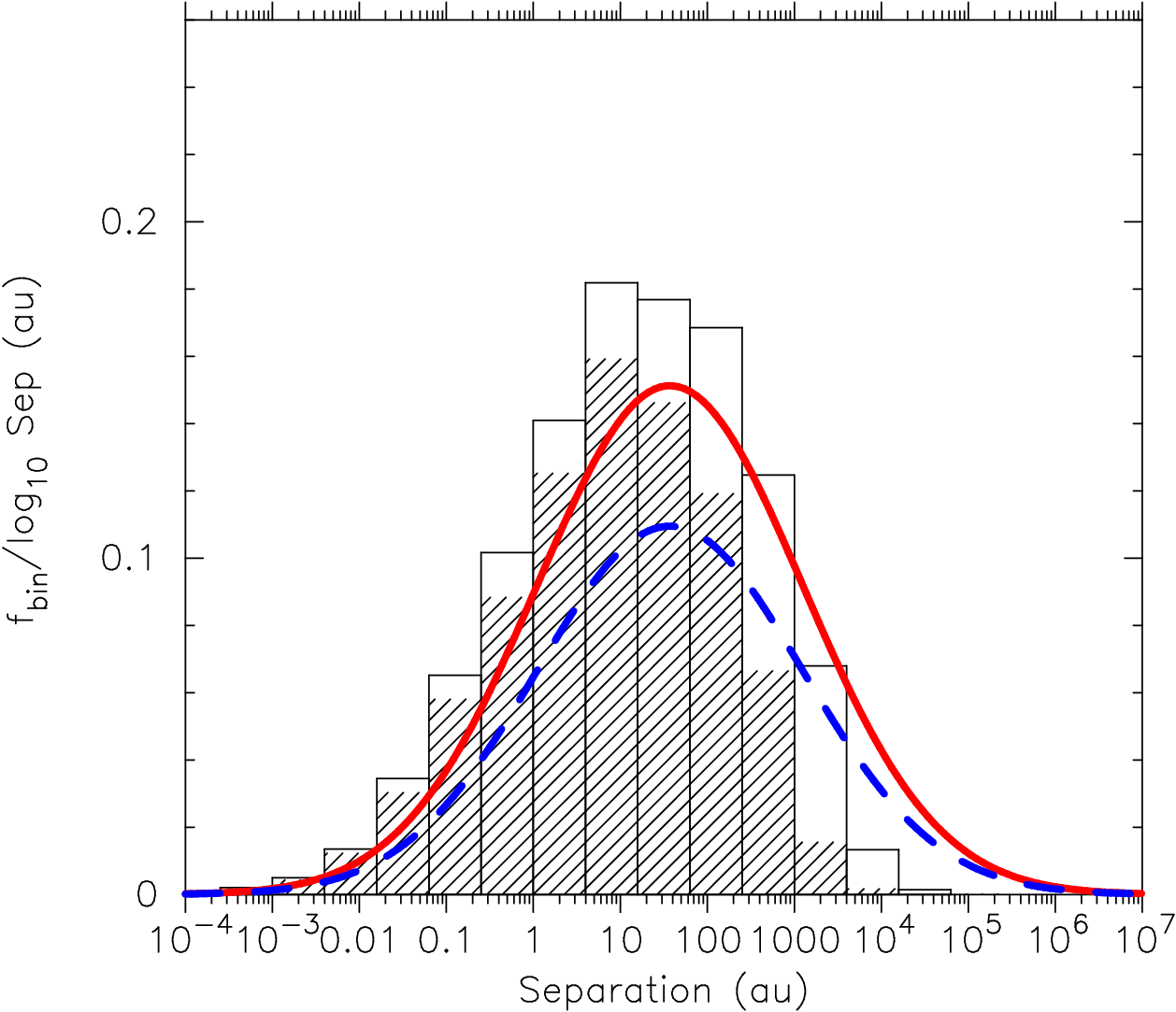}}}
  \end{center}
  \caption[bf]{The full separation distributions at 0\,Myr (open histograms) and 1\,Myr (hashed histograms) for clusters with an initial fractal dimension $D = 2.0$. Four different primordial binary population 
  set-ups are presented, and the log$_{10}$-normal fits to the separation distributions for G-dwarfs and M-dwarfs in the field (the (red) solid and (blue) dashed lines, respectively) are shown for comparison.}
  \label{sepdist2p0}
\end{figure*}

We evolve clusters with two different initial separation distributions. We consider clusters with separations drawn from the log$_{10}$-normal distribution observed for main sequence binaries in the field \citep{Duquennoy91,Raghavan10}, 
and also the inferred pre-main sequence distribution in \citet{Kroupa95a}. Three out of four clusters have the  DM91 separation distribution with varying primordial binary fractions; 100 per cent, 75 per cent 
and field-like; whereas the final cluster has the K95 separation distribution. The initial separation distributions (the open histograms), and the distributions after 1\,Myr (the hashed histograms) 
are shown in Fig.~\ref{sepdist2p0}. For comparison we show the log$_{10}$-normal fits to the separation distributions of field G-dwarfs (the (red) solid line), and field M-dwarfs (the (blue) dashed line). 

From inspection, we see that the results are similar to those obtained with the virialised, dense Plummer sphere models presented in \citet[][see their figs.~2~and~3]{Parker09a}. In the model in which we use the field separation 
distribution and binary fraction as our initial conditions, a significant amount of dynamical processing reduces the number of intermediate binaries, leading to an overall deficit of systems compared to the field. As found by 
\citet{Kroupa99}, the K95 separation distribution is reduced by interactions to the extent that the resultant separation distribution resembles that of the field for close and intermediate separation binaries. 

However, as noted by \citet{Parker09a}, 
and subsequent authors \citep{Kouwenhoven10,Moeckel10}, no cluster undergoing a dense phase will preserve the wide binary systems observed in the field, and other mechanisms are required to explain such systems 
\citep{Kouwenhoven10,Moeckel10,Moeckel11}.

\subsection{Visual binaries in the ONC}

\begin{figure*}
 \begin{center}
\setlength{\subfigcapskip}{10pt} \subfigure[100 per cent binary fraction, DM91 separation distribution]{\label{reipurth1p6-a}\rotatebox{270}{\includegraphics[scale=0.4]{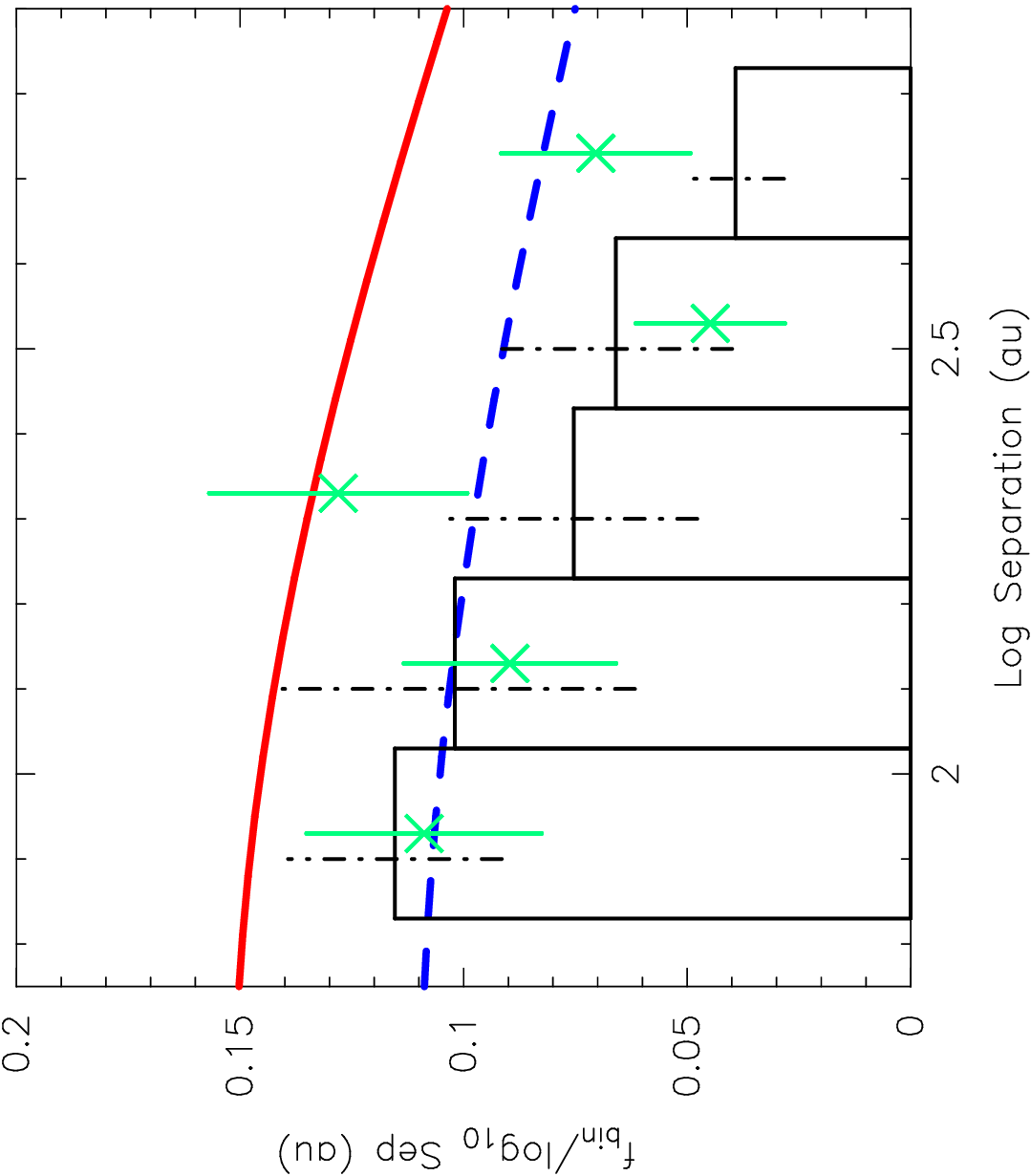}}}
\hspace*{0.6cm} 
\subfigure[100 per cent binary fraction, K95 separation distribution]{\label{reipurth1p6-b}\rotatebox{270}{\includegraphics [scale=0.4]{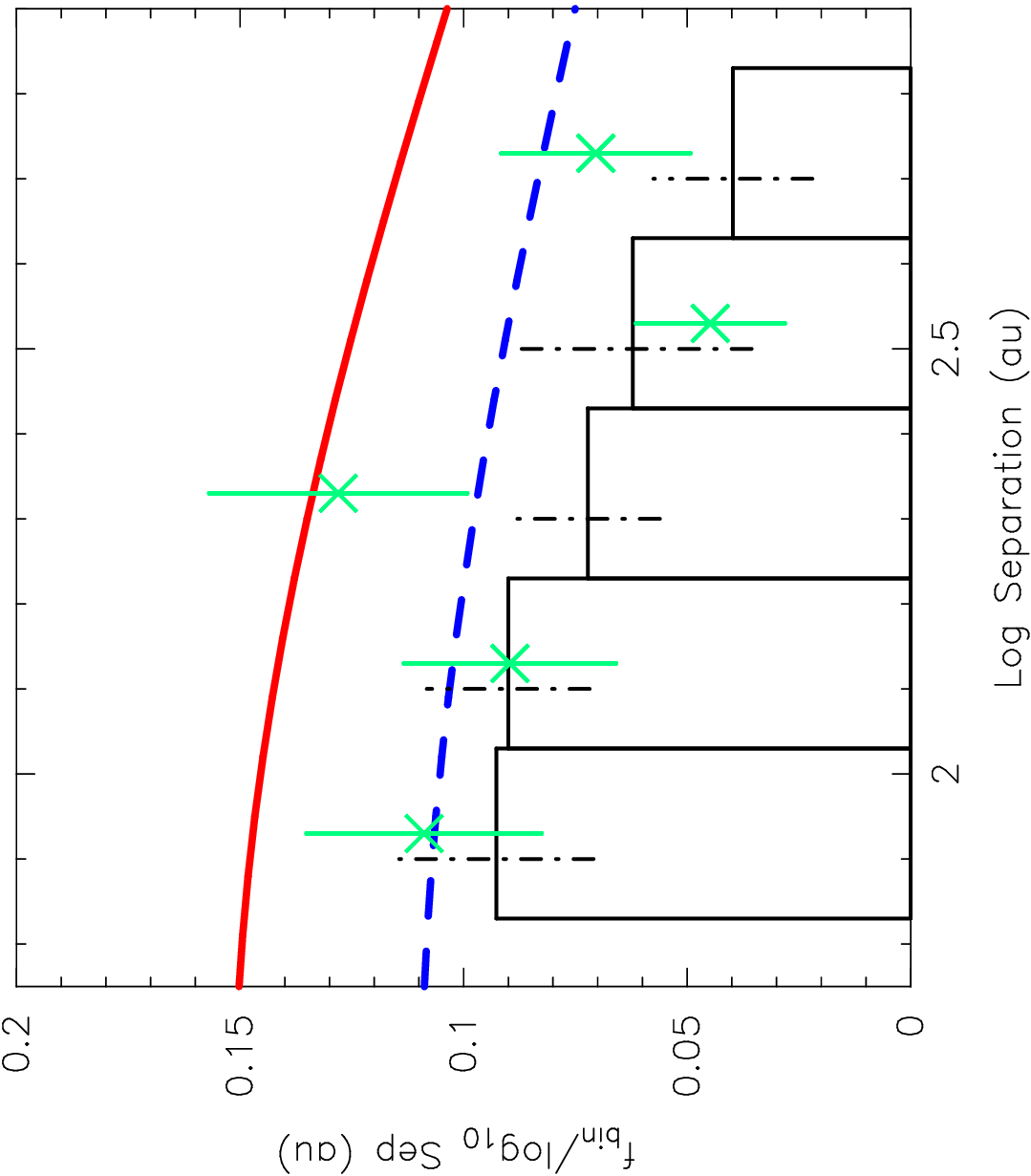}}}
\vspace*{0.5cm} \subfigure[Field binary fraction, DM91 separation distribution]{\label{reipurth1p6-c}\rotatebox{270}{\includegraphics[scale=0.4]{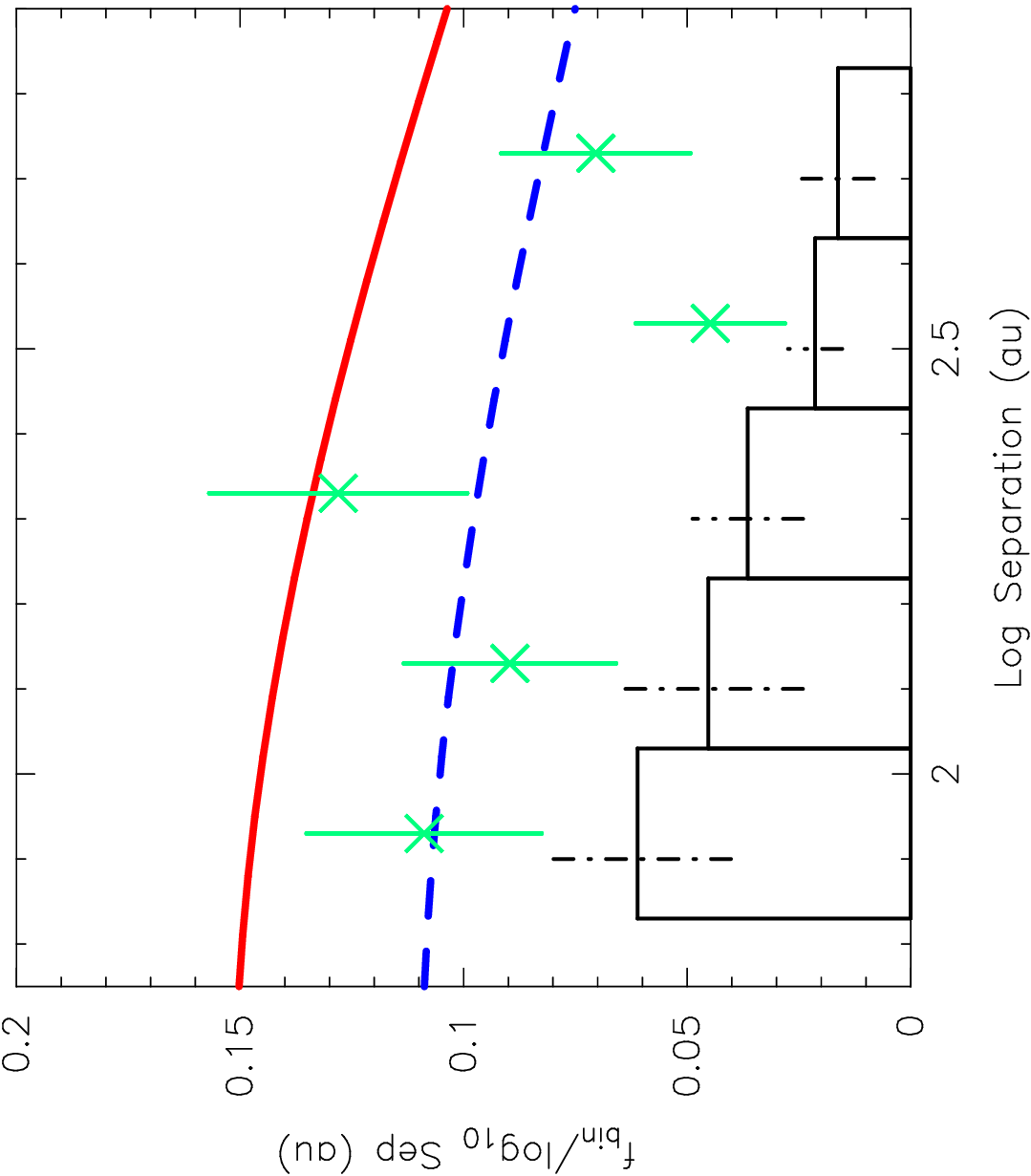}}}
\hspace*{0.6cm} \subfigure[75 per cent binary fraction, DM91 separation distribution]{\label{reipurth1p6-d}\rotatebox{270}{\includegraphics[scale=0.4]{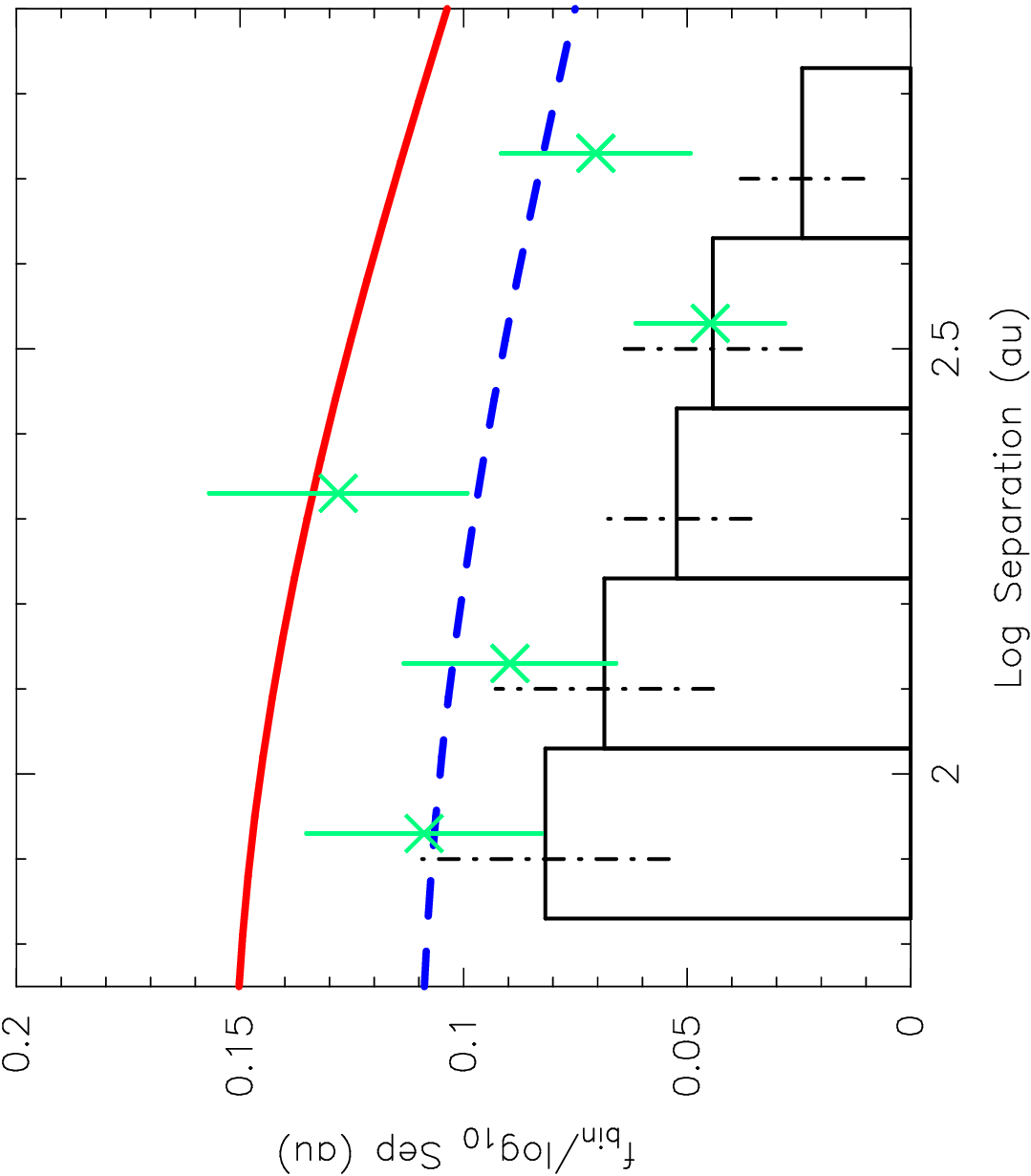}}}
  \end{center}
  \caption[bf]{Comparison with  the data for visual binaries in the ONC from \citet{Reipurth07}. Reipurth et al.'s data are shown by the (green) crosses with corresponding error bars. The separation 
  distribution, normalised to the binary fraction at 1\,Myr, from our simulations in clusters with an initial fractal dimension $D = 1.6$, are shown by the histograms. Four different primordial binary population 
  set-ups are presented, and the log$_{10}$-normal fits to the separation distributions for G-dwarfs and M-dwarfs in the field (the (red) solid and (blue) dashed lines, respectively) are shown for comparison.}
  \label{reipurth1p6}
\end{figure*}

\begin{figure*}
  \begin{center}
\setlength{\subfigcapskip}{10pt} \subfigure[100 per cent binary fraction, DM91 separation distribution]{\label{reipurth2p0-a}\rotatebox{270}{\includegraphics[scale=0.4]{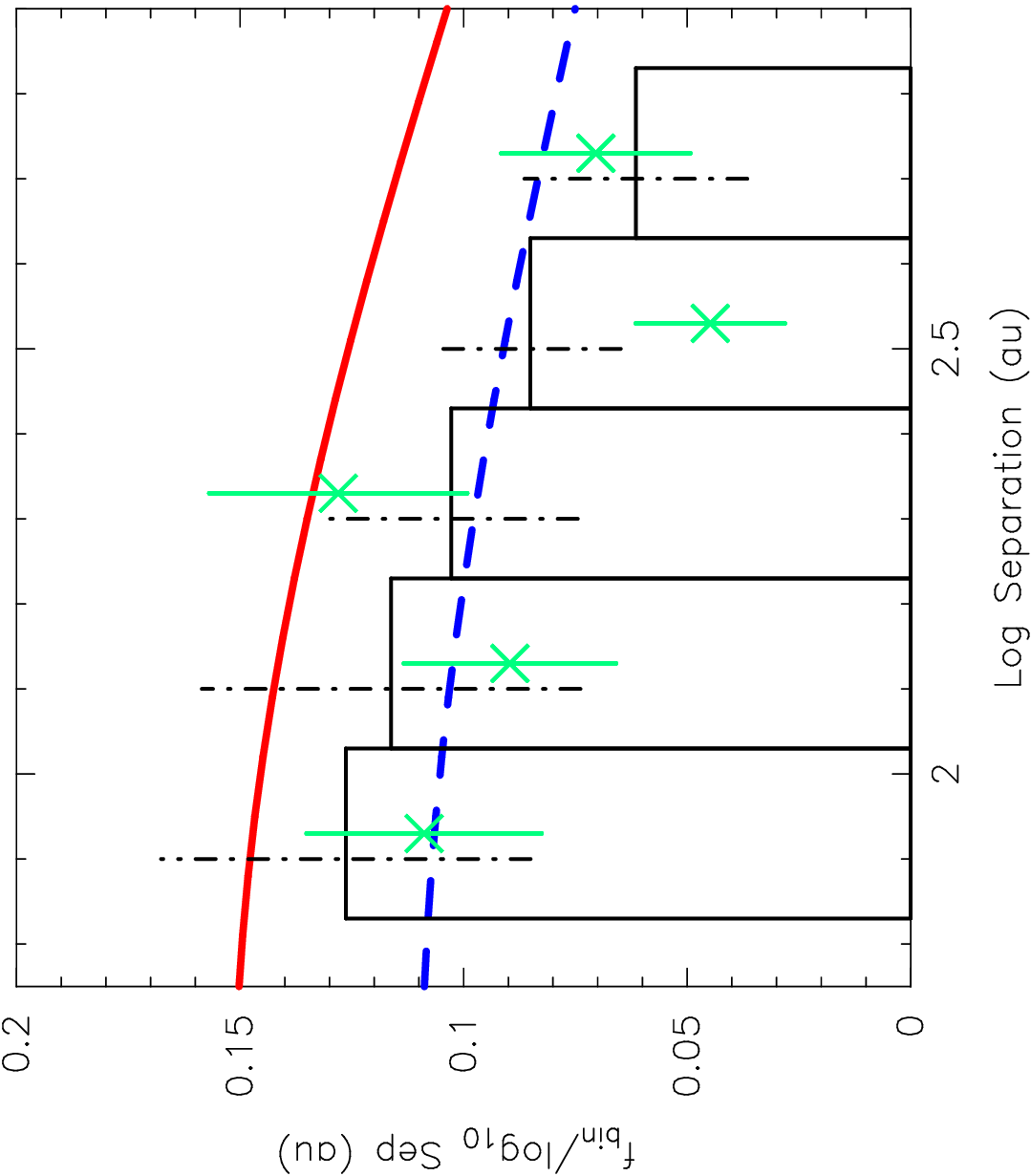}}}
\hspace*{0.6cm} 
\subfigure[100 per cent binary fraction, K95 separation distribution]{\label{reipurth2p0-b}\rotatebox{270}{\includegraphics [scale=0.4]{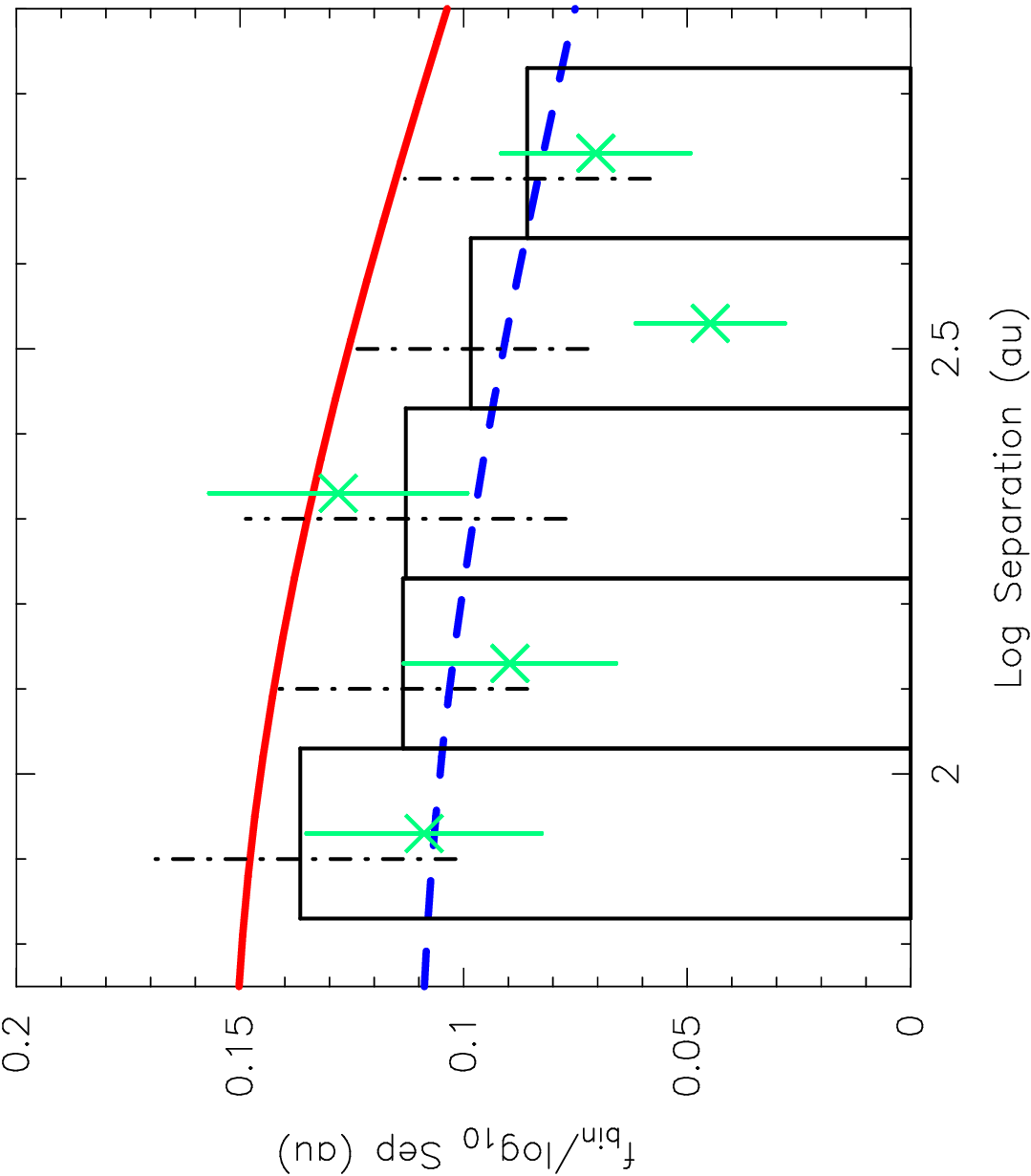}}}
\vspace*{0.5cm} \subfigure[Field binary fraction, DM91 separation distribution]{\label{reipurth2p0-c}\rotatebox{270}{\includegraphics[scale=0.4]{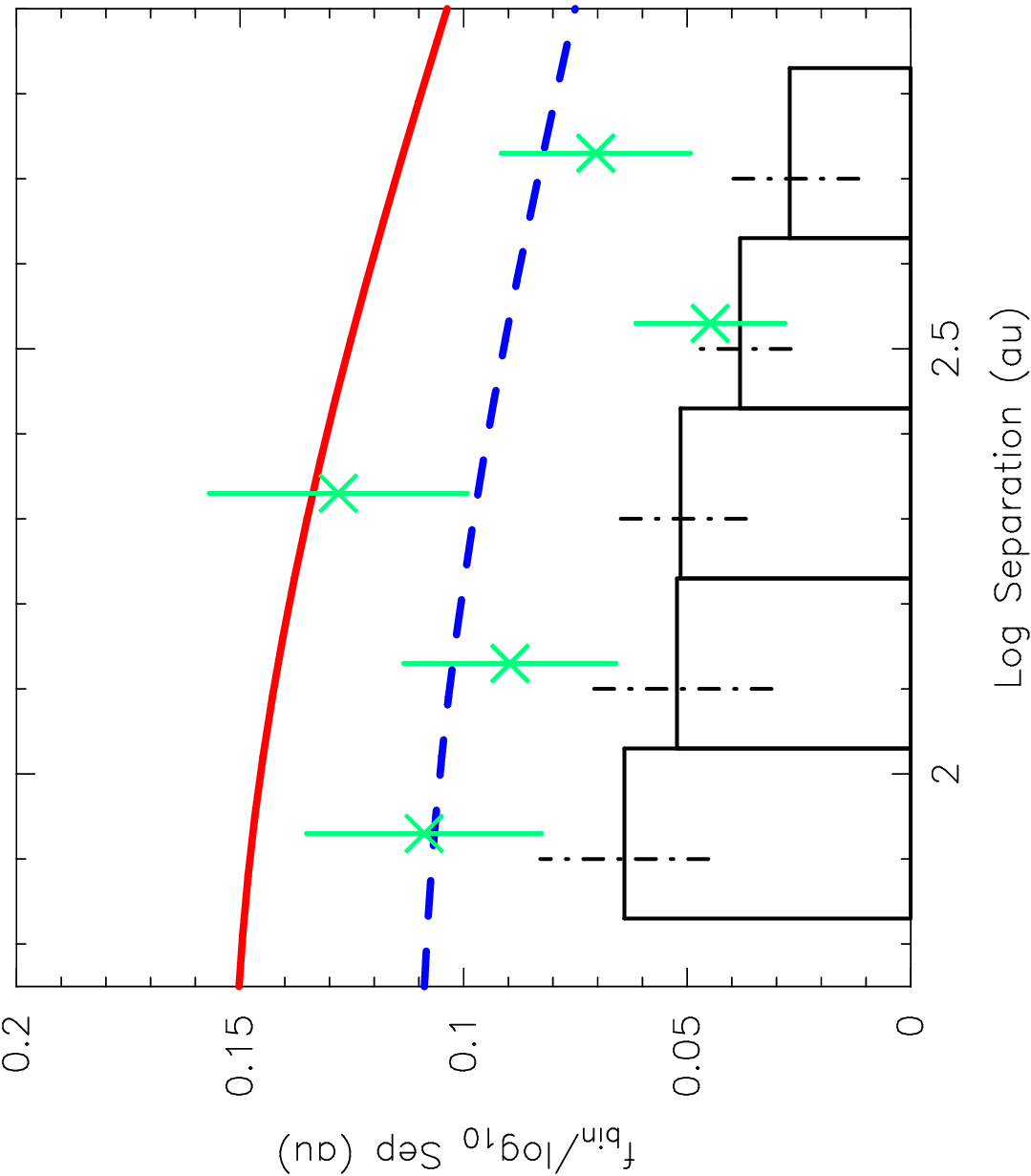}}}
\hspace*{0.6cm} \subfigure[75 per cent binary fraction, DM91 separation distribution]{\label{reipurth2p0-d}\rotatebox{270}{\includegraphics[scale=0.4]{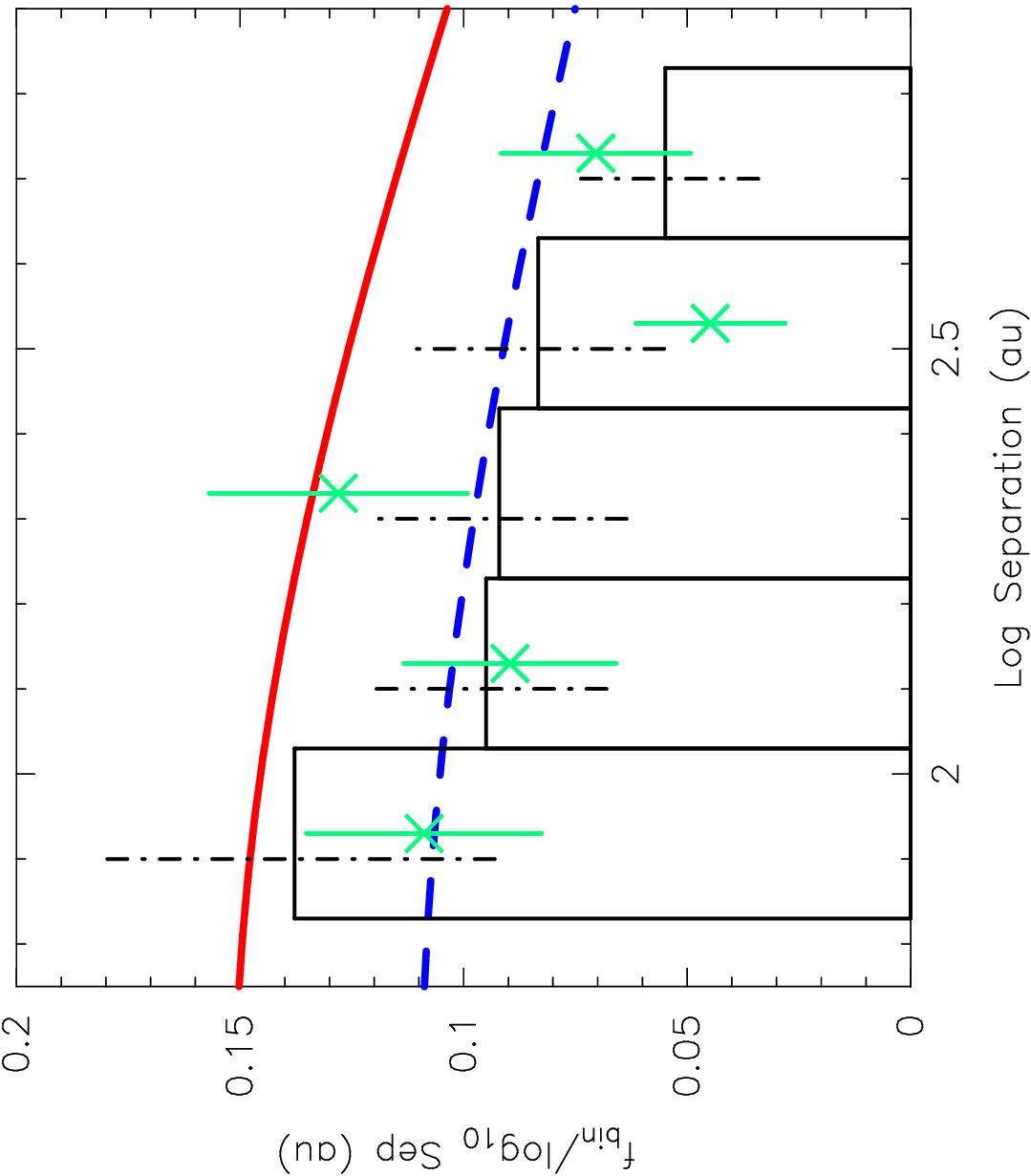}}}
  \end{center}
  \caption[bf]{Comparison with  the data for visual binaries in the ONC from \citet{Reipurth07}. Reipurth et al.'s data are shown by the (green) crosses with corresponding error bars. The separation 
  distribution, normalised to the binary fraction at 1\,Myr, from our simulations in clusters with an initial fractal dimension $D = 2.0$, are shown by the histograms. Four different primordial binary population 
  set-ups are presented, and the log$_{10}$-normal fits to the separation distributions for G-dwarfs and M-dwarfs in the field (the (red) solid and (blue) dashed lines, respectively) are shown for comparison.}
  \label{reipurth2p0}
\end{figure*}

\begin{figure*}
  \begin{center}
\setlength{\subfigcapskip}{10pt} \subfigure[100 per cent binary fraction, DM91 separation distribution]{\label{reipurth3p0-a}\rotatebox{270}{\includegraphics[scale=0.4]{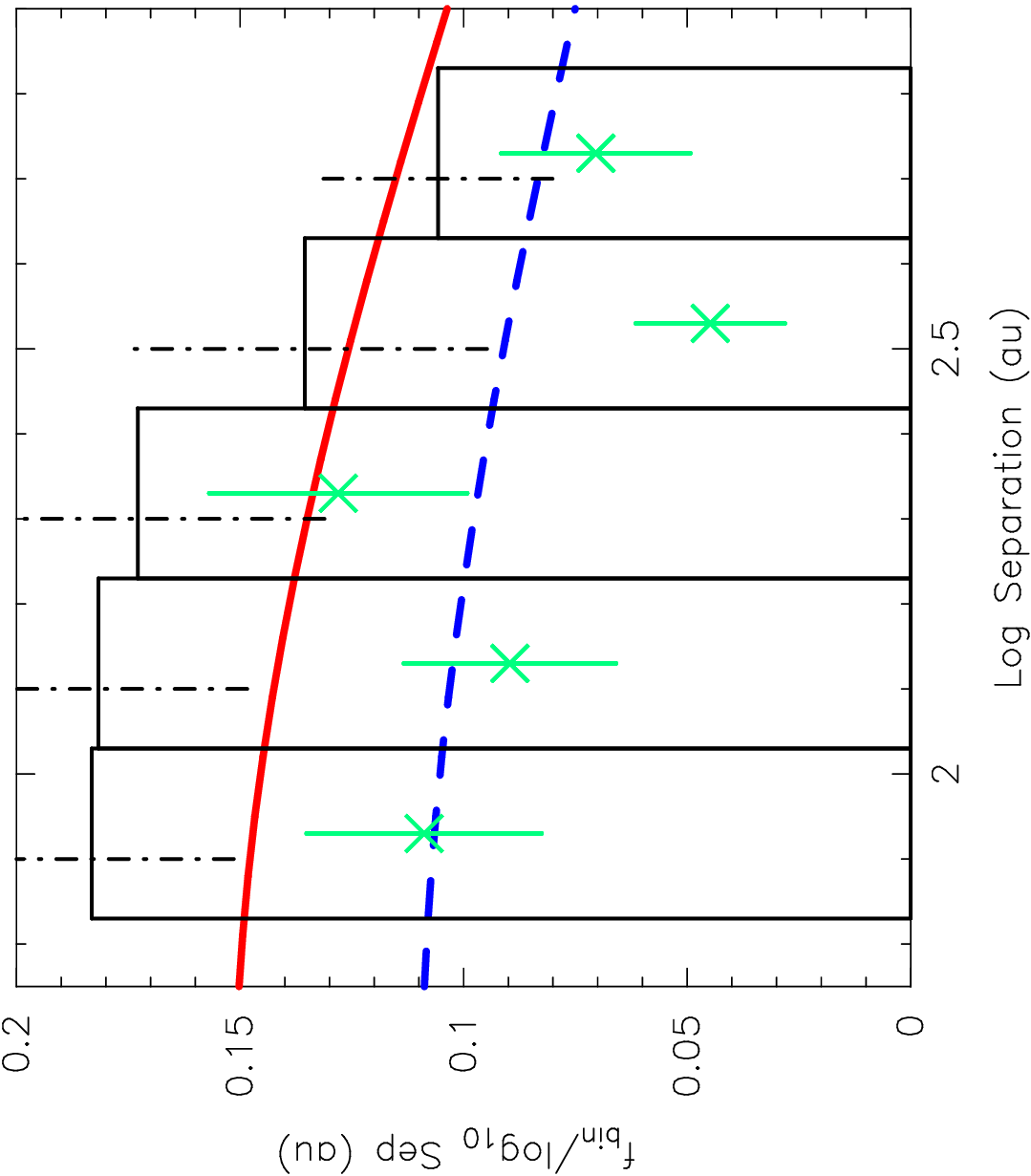}}}
\hspace*{0.6cm} 
\subfigure[100 per cent binary fraction, K95 separation distribution]{\label{reipurth3p0-b}\rotatebox{270}{\includegraphics [scale=0.4]{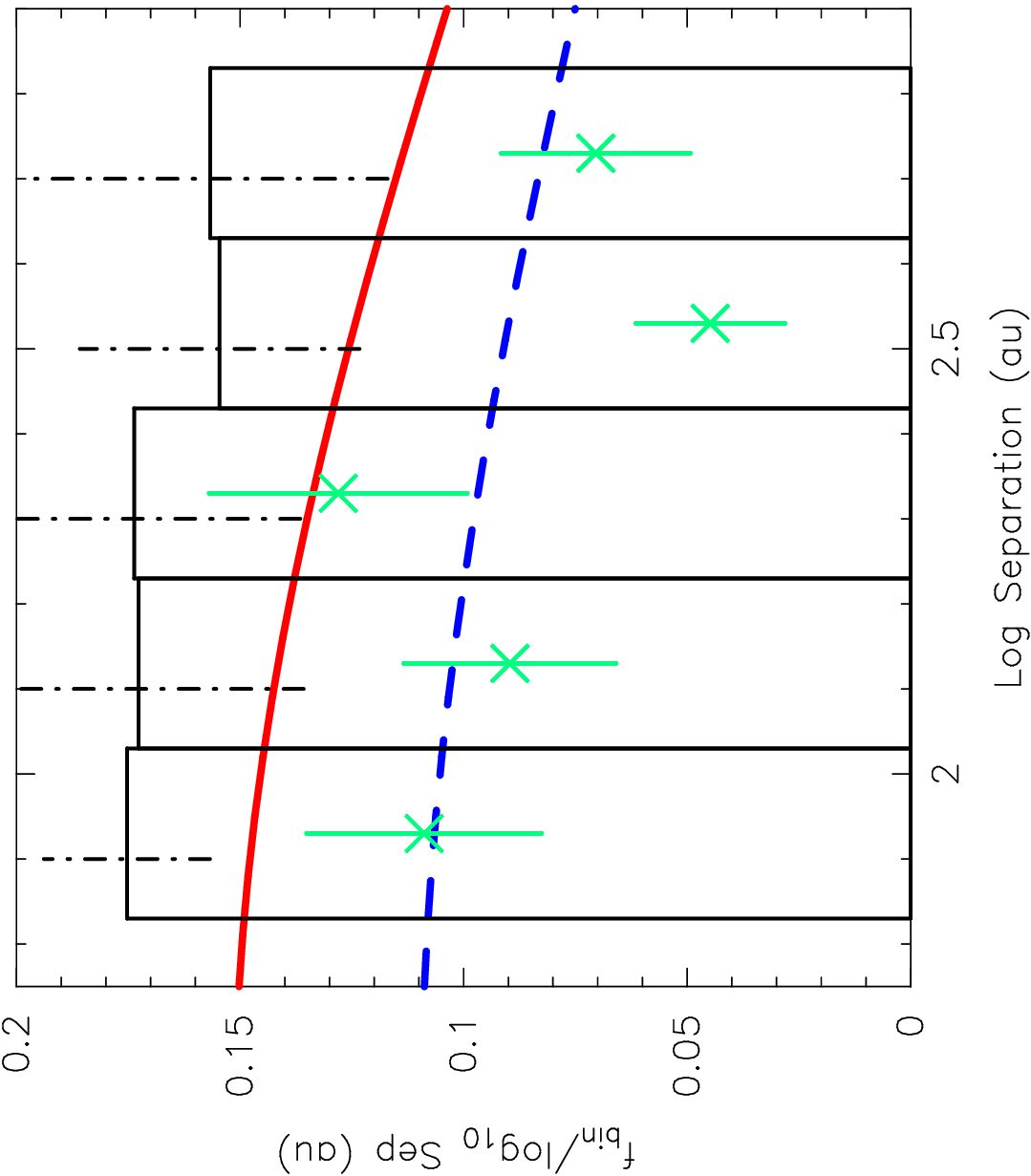}}}
\vspace*{0.5cm} \subfigure[Field binary fraction, DM91 separation distribution]{\label{reipurth3p0-c}\rotatebox{270}{\includegraphics[scale=0.4]{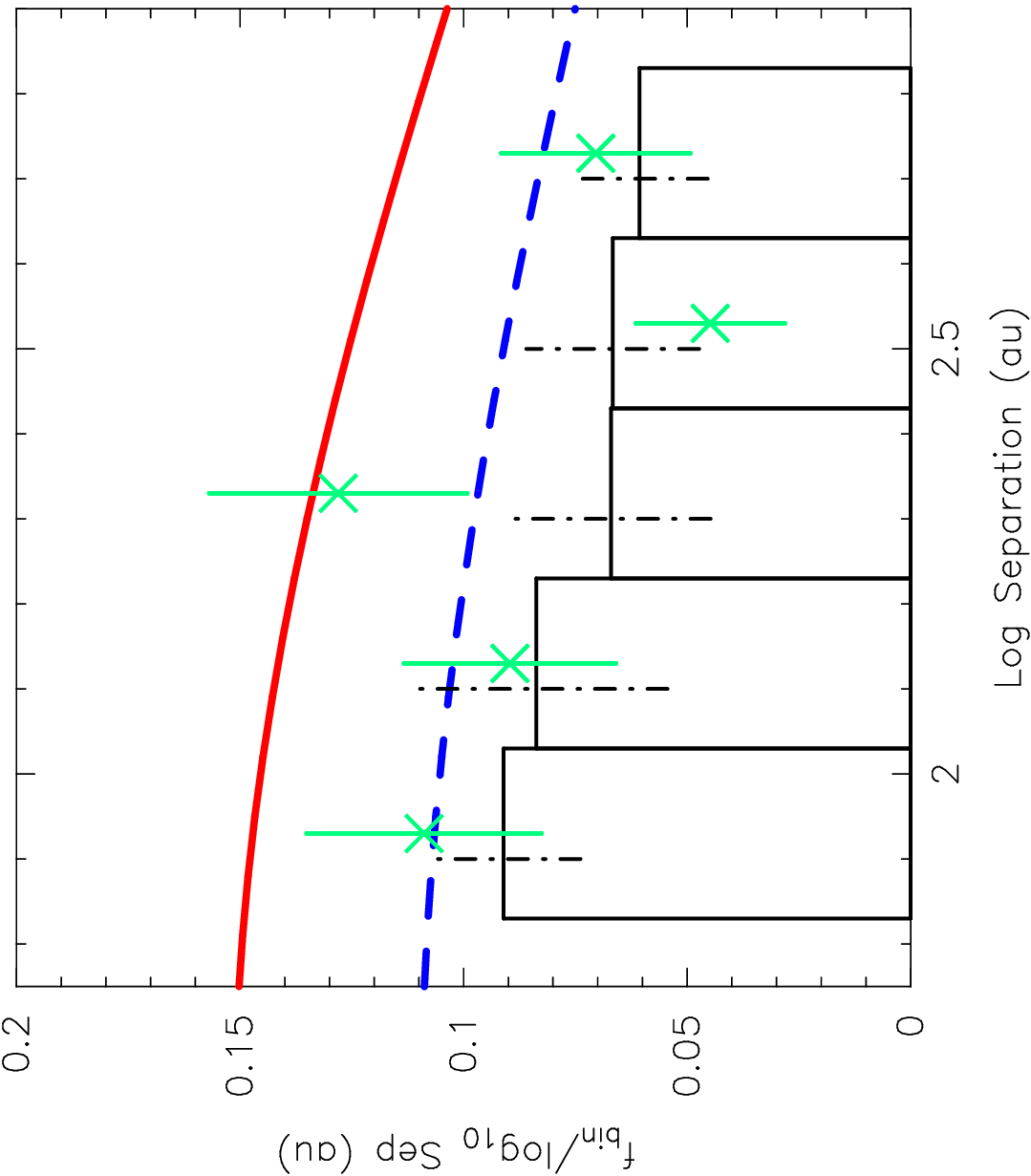}}}
\hspace*{0.6cm} \subfigure[75 per cent binary fraction, DM91 separation distribution]{\label{reipurth3p0-d}\rotatebox{270}{\includegraphics[scale=0.4]{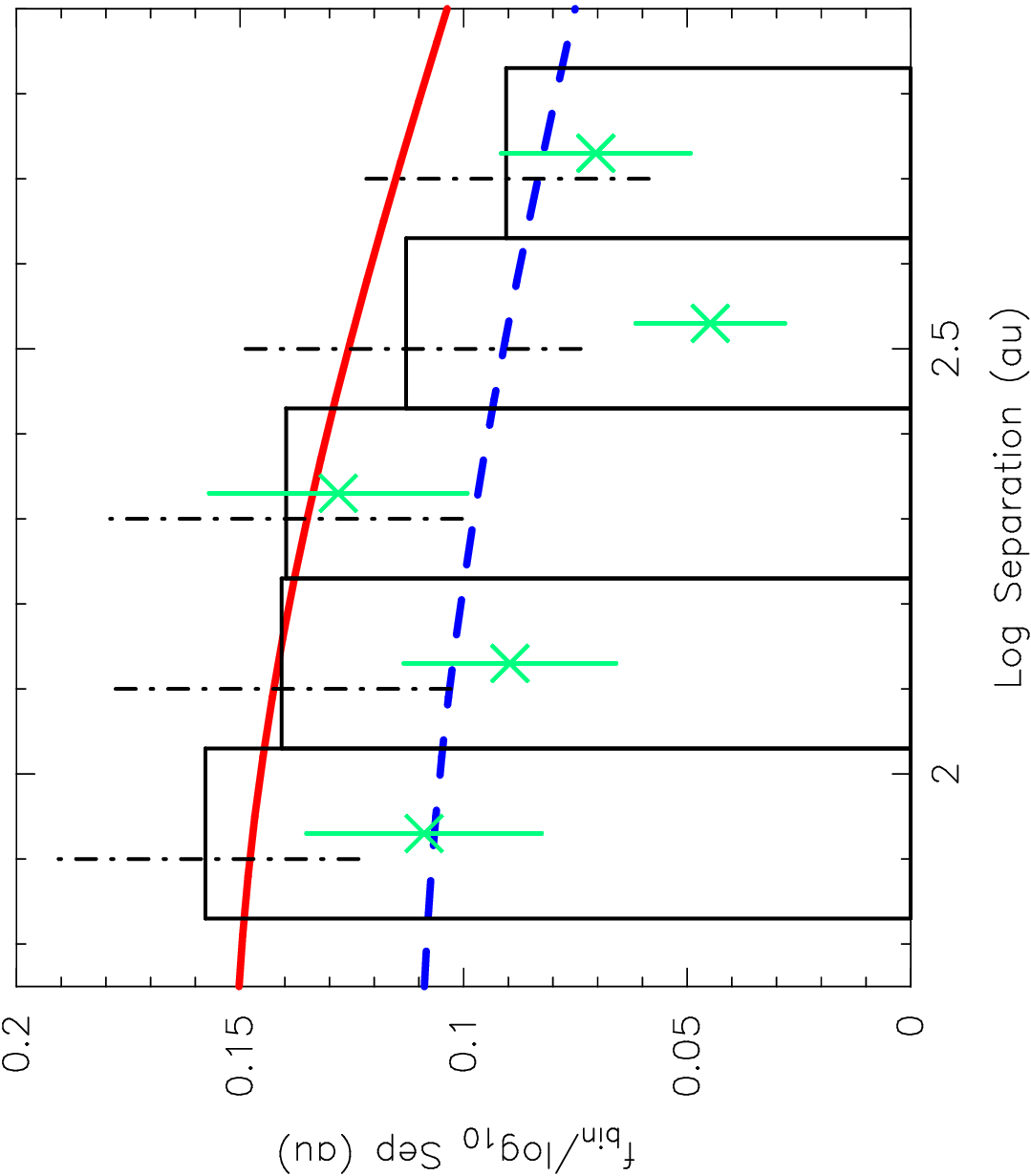}}}
  \end{center}
  \caption[bf]{Comparison with  the data for visual binaries in the ONC from \citet{Reipurth07}. Reipurth et al.'s data are shown by the (green) crosses with corresponding error bars. The separation 
  distribution, normalised to the binary fraction at 1\,Myr, from our simulations in clusters with an initial fractal dimension $D = 3.0$, are shown by the histograms. Four different primordial binary population 
  set-ups are presented, and the log$_{10}$-normal fits to the separation distributions for G-dwarfs and M-dwarfs in the field (the (red) solid and (blue) dashed lines, respectively) are shown for comparison.}
  \label{reipurth3p0}
\end{figure*}

Furthermore, we note that the field binary population is probably the sum of many differing star formation regions \citep{Brandner98,Goodwin10}, not all of which will have undergone the cool-collapse scenario presented here. Additionally, 
the most recent and complete observational census of binary systems in the ONC by \citet{Reipurth07} only considers visual binaries with separations in the range 67 -- 670\,au. For this reason, it makes more sense to compare the 
results of our simulations to the data from Reipurth et al., rather than the field separation distribution. The data from \citet{Reipurth07} are shown by the (green) crosses in Figs.~\ref{reipurth1p6},~\ref{reipurth2p0}~and~\ref{reipurth3p0}. 
We show our separation distributions (in the same range as \citet{Reipurth07}) after 1\,Myr of dynamical evolution with the histograms (and the corresponding error bars from averaging together 10 simulations for each plot).  

In Figs.~\ref{reipurth1p6},~\ref{reipurth2p0}~and~\ref{reipurth3p0}, we show the effects of dynamical evolution on the visual binaries in our clusters for the four initial binary populations. Firstly, we note that an initially field-like 
population (panel (c) in the figures) underproduces the required number of binaries in this separation range, apart from the clusters with smooth initial conditions (Fig.~\ref{reipurth3p0-c}). However, this cannot be the primordial 
binary population of the ONC because the overall binary fraction is lower than is observed (the dot-dashed line in Fig.~\ref{binfrac3p0}). 

Secondly, and following on from this, the other separation distributions for binaries in initially smooth clusters show that the number of visual binaries is overproduced (Figs.~\ref{reipurth3p0-a},~\ref{reipurth3p0-b}~and~\ref{reipurth3p0-d}). 

Finally, we see from inspection of Figs.~\ref{reipurth1p6}~and~\ref{reipurth2p0} that \emph{all} populations with an initial binary fraction of either 75 per cent or 100 per cent reproduce the observed separation distribution within the uncertainties, 
suggesting that there must have been an overabundance of binaries with separations in this range at the birth of the cluster. Because of the highly uncertain binary fraction in the ONC \citep{Petr98,Kaczmarek11}, we see from inspection of 
Fig.~\ref{bin_fracs} that clusters with an initial fractal dimension of $D = 2.0$ or $1.6$ are equally consistent with the observations, assuming either a DM91 or K95 initial separation distribution, and a binary fraction between 0.75 and unity.

\section{Discussion}
\label{discuss}

We have examined the dynamical evolution of fractal clusters in cool collapse with three different levels of initial substructure. We consider clusters with fractal dimensions of 1.6 (very clumpy), 2.0 and 3.0 (a roughly uniform sphere). 
In each of these substructured clusters, we examine the effects of this cool collapse on four different primordial binary populations, which are characterised by the primordial binary fraction and binary separation distribution. 
Observations indicate that many young star forming regions are both substructured, and that stars form with subvirial velocities \citep[e.g.][]{Peretto06,Proszkow09}. Star clusters with these characteristics were used by \citet{Allison09b,Allison10} to show that mass segregation in the ONC can 
occur dynamically on a very short timescale (1\,Myr), negating the need for primordial mass segregation in the ONC \citep{Bonnell98}. Furthermore, \citet{Allison11} have shown that trapezium-like systems regularly form in 
such simulations, suggesting the cool-collapse of a clumpy cluster could be the most likely dynamical evolution scenario for the ONC. The most favourable initial conditions for this dynamical mass 
 segregation (and the formation of the Trapezium system) are a clumpy ($D \leq 2.0$), cool ($Q < 0.4$) cluster \citep{Allison09b,Allison10,Allison11}.

The hypothesis presented in \citet{Allison09b} is supported by observations. The outskirts of the ONC (20\,pc from the Trapezium) appear to be subvirial and in cool collapse \citep{Feigelson05,Tobin09}, whereas the 
velocity dispersion in the centre is 4.3\,kms$^{-1}$, much higher than the value we would expect if the ONC was in virial equilibrium \citep[2.5\,kms$^{-1}$ --][]{Olczak08}. This suggests that the centre of the ONC has already 
undergone cool-collapse, and is now expanding. 

However, for simplicity \citet{Allison09b} did not include primordial binaries in their simulations. The binary fraction in the ONC is not negligible, and is consistent with the field value 
\citep[between 40 and 60 per cent][]{Petr98,Reipurth07}. Several authors \citep{Kroupa99,Parker09a,Kaczmarek11} have proposed that the ONC was born with a much higher binary fraction than its present value, and that 
dynamical interactions have processed this primordial population to that which we observe today. For this to happen, the cluster must have undergone a dense phase during its evolution. If the results of dynamical processing on 
a primordial binary population can be reconciled with the observed binary fraction and separation distribution, then this provides strong support for such a theory. However, it is unclear how a significant number of binaries 
could form in such a dense environment \citep{Bate09,Parker09a,Moeckel10}. An alternative scenario is that the initial density of the cluster is such that binary formation is not impeded, but the cluster then undergoes a dense phase 
via the collapse of a substructured fractal \citep{Allison09b,Allison10}. An excellent test of this hypothesis is to study the effects of cool-collapse on a primordial binary population.

The low, field-like binary fraction in Orion is in itself not conclusive proof that the cluster has undergone significant dynamical processing; a more stringent test is to examine the separation distribution of the cluster. \citet{Reipurth07} 
conducted a survey of visual binaries in the ONC, corresponding to a separation range 67 -- 670\,au. If the ONC did go through a dense phase, then we would expect the hard-soft boundary for binary disruption 
\citep{Heggie75,Hills75a} to lie within this range \citep{Parker09a}. Therefore, by examining the effects of cluster evolution on various primordial binary populations, we can constrain the primordial binary fraction \emph{and} 
separation distribution (in this separation range) based on comparison with the \citet{Reipurth07} data.

Direct comparison of the observations with our simulations is presented for each initial level of subclustering in Figs.~\ref{reipurth1p6},~\ref{reipurth2p0}~and~\ref{reipurth3p0}. Firstly, we note that the process of cool-collapse in 
clusters can reproduce the observed separation distribution to zeroth order for most primordial binary populations. Clusters with a moderate to high level of subclustering cannot preserve enough binaries in the separation range 
67 -- 670\,au for an initially field-like binary fraction (Figs.~\ref{reipurth2p0-c}~and~\ref{reipurth1p6-c}). If we start the cluster as a uniform sphere ($D = 3.0$), then it is possible to reproduce the observations with a field-like binary 
fraction and separation distribution  (Fig.~\ref{reipurth3p0-c}). However, the overall binary fraction for the cluster is still too low (Fig.~\ref{binfrac3p0}), suggesting that even in this more placid dynamical scenario, the primordial binary 
fraction has to be larger than the present day. Furthermore, all other initial binary populations do not undergo enough processing to suggest that this fractal dimension is a realistic initial condition for the ONC. 


A moderate level of of substructure ($D = 2.0$) results in excellent agreement with the observations of \citet{Reipurth07} for clusters with DM91 separation distributions and primordial binary fractions of 100 or 75 per cent 
(Figs.~\ref{reipurth2p0-a}~and~\ref{reipurth2p0-d}, respectively). 
In clusters with very clumpy initial conditions ($D = 1.6$),  
the level of dynamical processing is too extreme in all but the cluster with a DM91 separation distribution and a 100 per cent primordial binary fraction (Fig.~\ref{reipurth1p6-a}) to be reconciled 
with the observations of \citet{Reipurth07}. 


As discussed in \citet{Allison10} and \citet{Allison11}, dynamical mass segregation and the formation of trapezium systems can be very transient. In order to reproduce the observed level of mass segregation it is favourable to 
have clumpy initial conditions. If we assume that clumpy, cool initial conditions are required for 
the ONC to mass segregate and form the Trapezium system, then the observed binary fraction and separation distribution requires an initially higher binary fraction ($\sim$ 70 -- 80 per cent) than is observed today. 

Finally, we note that if the clusters are initially clumpy, the majority of binaries are processed before the cluster reaches its densest phase during the collapse. This is because the pockets of substructure are dense enough initially to affect the binaries, 
and suggests that all star clusters that form with substructure will process a primordial binary population, irrespective of whether the cluster undergoes cool-collapse (which exacerbates the processing), remains in virial equilibrium, or expands. 

Therefore, even some expanding associations which form supervirial/unbound may not preserve their primordial binary populations. However, we note that the fractals we set up have initial densities of $\sim$~300\,M$_\odot$pc$^{-3}$ (see Fig.~\ref{cluster_evolve}), which are higher than many star forming regions that 
will subsequently become unbound associations \citep[e.g.][]{Jorgensen08,Gutermuth09,Bressert10}. We will further investigate the effects of dynamical evolution on such sparse regions in a future paper. 

\section{Conclusions}
\label{conclude}

We present the results of $N$-body simulations of fractal star clusters containing $N = 1500$ stars in cool-collapse, in order to investigate the effect of this dynamical evolution scenario on various primordial binary populations. We have varied 
the initial level of substructure in the cluster, the primordial binary fraction, and the initial separation distribution.  Our conclusions can be summarised as follows:

(i) Primordial binary populations are heavily processed in clusters undergoing cool-collapse. Qualitatively, the results are similar to those from dynamical evolution of the binary population in initially very dense virialised Plummer 
 spheres \citep{Kroupa99, Parker09a}. 
 
(ii) The level of dynamical processing varies as a function of the fractal dimension; clumpy clusters break up more binaries than smoother clusters.

(iii) The majority of dynamical processing in substructured clusters occurs before the cluster reaches its densest phase; therefore, it is the initial densities in the substructure which is the most significant contributor to 
altering the binary population, rather than the cool-collapse itself. This suggests that even some star-forming regions that do not collapse will significantly process a primordial binary population.

(iv) If clusters undergo cool-collapse, then the field binary fraction and separation distribution cannot be the primordial distribution in the ONC. Comparison of our simulations with observations suggests that the ONC had a primordial binary fraction of between 75 
 and 100 per cent.
 
 We demonstrate that the cool-collapse scenario, which is consistent with the filamentary, subvirial early phases of star formation, and can explain the level of mass segregation in the ONC through dynamics, also reproduces the observed binary fraction 
 and separation distribution. If a moderate to high level of substructure is required to produce dynamical mass segregation, then an $\sim$80 per cent binary fraction, and field-like separation distribution with a cut-off around $5 \times 10^3$\,au 
represents the most likely initial binary population.

\section*{Acknowledgements}

We thank the anonymous referee for their prompt report and helpful suggestions on the original draft. The simulations in this work were performed on the \texttt{BRUTUS} computing cluster at ETH Z{\"u}rich. RJA acknowledges support from the Alexander von Humboldt Foundation in the form of a research fellowship.

\bibliographystyle{mn2e}
\bibliography{frac_bin}

\label{lastpage}

\end{document}